# PepLand: a large-scale pre-trained peptide representation model for a comprehensive landscape of both canonical and non-canonical amino acids

Ruochi Zhang[1,2,3], Haoran Wu[3], Yuting Xiu[3], Kewei Li[1,4], Ningning Chen[3], Yu Wang[3], Yan Wang[1,2,4], Xin Gao[5,6,*], Fengfeng Zhou[1,4,*].

1 Key Laboratory of Symbolic Computation and Knowledge Engineering of Ministry of Education, Jilin University, Changchun, Jilin, China, 130012.

2 School of Artificial Intelligence, Jilin University, Changchun, China,130012.

3 Syneron Technology, Guangzhou, China, 510700.

4 College of Computer Science and Technology, Jilin University, Changchun, Jilin, China, 130012.

5 Computational Bioscience Research Center, King Abdullah University of Science and Technology (KAUST), Thuwal, Saudi Arabia, 23955.

6 Computer Science Program, Computer, Electrical and Mathematical Sciences and Engineering Division, King Abdullah University of Science and Technology (KAUST), Thuwal, Saudi Arabia, 23955.

* Correspondence may be addressed to Fengfeng Zhou: FengfengZhou@gmail.com or ffzhou@jlu.edu.cn . Lab web site: http://www.healthinformaticslab.org/. Phone: +86-431-8516-6024. Fax: +86-431-8516-6024. Correspondence may also be addressed to Xin Gao: xin.gao@kaust.edu.sa.

# Abstract

In recent years, the scientific community has become increasingly interested on peptides with non-canonical amino acids due to their superior stability and resistance to proteolytic degradation. These peptides present promising modifications to biological, pharmacological, and physiochemical attributes in both endogenous and engineered peptides. Notwithstanding their considerable advantages, the scientific community exhibits a conspicuous absence of an effective pre-trained model adept at distilling feature representations from such complex peptide sequences. We herein propose PepLand, a novel pre-training architecture for representation and property analysis of peptides spanning both canonical and non-canonical amino acids. In essence, PepLand leverages a comprehensive multi-view heterogeneous graph neural network tailored to unveil the subtle structural representations of peptides. Empirical validations underscore PepLand's effectiveness across an array of peptide property predictions, encompassing protein-protein interactions, permeability, solubility, and synthesizability. The rigorous evaluation confirms PepLand's unparalleled capability in capturing salient synthetic peptide features, thereby laying a robust foundation for transformative advances in peptide-centric research domains. We have made all the source code utilized in this study publicly accessible via GitHub at https://github.com/zhangruochi/pepland

# Introduction

Peptides, being naturally occurring biological molecules, have increasingly become the cornerstone in drug development due to their inherent therapeutic properties [1]. Their high specificity, potent efficacy, and relative safety contribute significantly towards their growing appeal in the pharmaceutical landscape [2]. Non-canonical amino acids serve as potent foundational elements for the development of peptide-based materials and therapeutic agents, thereby facilitating chemists in transcending the structural and functional constraints traditionally associated with the limited repertoire of canonical amino acids [3]. For instance, cyclic peptides are chains of amino acids, both canonical and non-canonical, that are linked at distant points to create macrocyclic structures [4]. Cyclic peptides exhibit diverse biological activities, from acting as signaling agents in complex biological processes to serving as chemical weapons for defense, highlighting the vast functional potential of cyclic peptides [5].

Deep learning techniques, particularly in the prediction of peptide properties and peptide-protein interaction models, are being extensively explored to further accelerate the pace of drug discovery and development processes [6 7]. However, the application of these models presents unique challenges. Pre-existing models, such as the Evolutionary Scale Modeling (ESM) [8] and ProteinBert [9], leverage amino acid sequences to learn and predict coevolutionary information embedded in protein sequences. These models have showcased noteworthy success in protein-related tasks, but their efficacy is compromised when dealing with peptides. This shortcoming arises from the fundamental differences between proteins and peptides. Peptides generally have shorter lengths compared to proteins, hence their data distribution does not align well with the existing models that have been trained predominantly on protein sequences. Furthermore, a critical limitation observed in models such as ESM is their incapacity to effectively handle non-canonical amino acids, which are frequently used to enhance the pharmaceutical properties of peptides [9 10]. There are some researchers who might use models from the field of ligands(small molecules) to extract the representation of peptides[11], but our results also indicate that this is not a suitable approach. Essentially, peptides and small molecules have different data distributions, such as the fact that the length of peptide SMILES(simplified molecular-input line-entry system) [12] is usually 5 to 10 times longer than that of small molecules.

To address this existing gap, we have meticulously devised a fragmentation algorithm tailored specifically for the structural analysis of peptides containing non-canonical amino acids, thereby facilitating their breakdown into optimally granular components. Our investigations reveal the paramount importance of granularity in accurately predicting the resultant properties of these peptides. Concurrently, we introduce PepLand, an innovative pre-training model equipped with the distinctive capability to simultaneously process both canonical and non-canonical amino acids, thereby significantly amplifying its versatility and efficacy in peptide-centric endeavors. Our model ingeniously amalgamates a multi-view heterogeneous graph that comprises both atom and fragment views, thus enabling a comprehensive representation of peptides at varying granular levels. Additionally, we implement a two-step training strategy to enable the model to efficiently learn from data related to both canonical and non-canonical amino acids, in spite of the latter's smaller dataset size. The experimental results show that PepLand achieves state-of-the-art results in various peptide property prediction tasks, such as protein-peptide interaction, membrane permeability, solubility, etc. We have also conducted rigorous experiments to demonstrate that the granularity of fragments is crucial for the final property prediction task. Due to the immense potential value of cyclic peptides, we have demonstrated in a case study that PepLand has a high accuracy in predicting the affinity of cyclic peptides.

We firmly believe that our model, PepLand, signifies a major step forward in the utilization of deep learning models in the realm of peptide-based drug development. By facilitating a comprehensive understanding of peptide properties, including those incorporating non-canonical amino acids, PepLand stands to revolutionize the field of peptide representation learning, paving the way for the accelerated discovery of more effective therapeutic peptide candidates.

# Material and Method

## Training dataset

The training set is a mixture of three sub-datasets: For the first stage of pre-training, we utilized a protein dataset consisting of both canonical and non-canonical amino acids, comprising a total of 7,924,509 samples. Most of the data were extracted from UniProt [13], which is a comprehensive and widely used resource in the field of bioinformatics. In order to meet our requirements, we have filtered out the sequence more than 30 in length, resulting in a subdataset of UniPort with a total data volume of 8 million entries.

For the second pre-training stage, another dataset involved is CycPeptMPDB [14] which consists of 7334 cell-penetrating peptides with non-canonical amino acids. In addition, some high-quality data from PDB [15] which contains non-canonical amino acids, were incorporated into the dataset. This final dataset contained a total of 8,977 peptide sequences which all contains non-canonical amino acids.

## Evaluating dataset

To verify the effectiveness of the pre-trained model, we constructed a collection of evaluating datasets related to peptides property prediction. These datasets include transmembrane prediction, solubility prediction, and protein-peptide affinity prediction tasks. We created tasks and datasets for peptides containing only canonical amino acids as well as peptides containing non-canonical amino acids.

Canonical CPP dataset : We evaluate our models on the task of cell penetrating ability prediction on canonical peptide sequences. Our CPP dataset is collected from 22 relevant cell-penetrating peptide databases by compiling literature on existing cell-penetrating peptide prediction models [16 16 17], containing 1162 positive and negative samples each. To ensure the reliability of the sources, we traced each database and listed the relationships between the databases. Besides, we also checked whether there were conflicts and duplicates in different databases, and the anomalous data were checked artificially again.

Non-canonical CPP dataset: CycPeptMPDB is the first web-accessible database of cyclic peptide membrane permeability. CycPeptMPDB has collected information on a total of 7334 cyclic peptides, including the structure and experimentally measured membrane permeability, from 45 published papers and 2 patents from pharmaceutical companies. Precisely, CycPeptMPDB contains two types of information: the first is physical quantities of membrane permeability such as LogP (an index of lipophilicity). The second is sequence information described by HELM and monomers as partial structures constituting the cyclic peptides. In contrast with the previous CPP dataset, CycPeptMPDB is suitable for the regression task of membrane permeability, yielding a refined estimation of cell-penetrating ability.

Solubility dataset: We utilize PROSO-II dataset [18] for our peptide solubility prediction task. Thanks to the restrictive data selection from the pepcDB and PDB databases, PROSO-II dataset is one of the largest available databases used for solubility-model building and evaluation. A large portion of the data comes from pepcDB database, in which each protein is associated with multiple amino acid sequences corresponding to different constructs. For each construct experimental result, its status history is recorded, including the one of solubility. Hence, all constructs that achieved the soluble status are considered as positive samples. Besides, additional soluble data originates also from PDB entries of heterologous complexes, the proteins with the annotation of 'Expression Organism: ESCHERICHIA COLI' are considered positive. Adapting PROSO-II dataset to our needs, we have constructed a subset (about 4 thousand pieces of data) which contains only sequences less than 50 in length.

Canonical Binding Affinity dataset: We construct our binding affinity prediction benchmark data set as the same workflow of [6] . Overall, we got a total of 1781 protein-peptide pairs with binding affinity.

Non-Canonical Binding Affinity Dataset: We also extracted a subset from the aforementioned Binding Affinity Prediction dataset, in which all peptides contain non-canonical amino acids.

## Overall Workflow for PepLand

In our study, we constructed a multi-view heterogeneous graph that encompasses three views, namely, the atom view, junction view and fragment view. Nodes and edges between these two views are allowed to exchange information through a message passing mechanism. In a heterogeneous graph, we use atoms as nodes and keys as edges. This forms our atom view. The atom view is designed to capture atomic-level representations of the peptides. Non-canonical amino acid modifications in peptides typically follow specific patterns. For example, common D-amino acids modifications can enhance peptide stability [19]. Similarly, α-amino acids modifications can enhance the biological stability and resistance to degradation of peptides [19]. Some non-canonical amino acid modifications of proline are promising candidates for conformational studies and for tuning the biological, pharmaceutical, or physicochemical properties of naturally occurring, as well as de novo designed, linear, and cyclic peptides [4,10,20]. Based on these characteristics, we have developed a fragment method for peptide representation. These fragments serve as another class of nodes in the heterogeneous graph, with bonds between fragments acting as edges in the graph. The fragment view implicitly captures the properties of non-canonical amino acids we want to learn.

During the training phase, we adopt an attribute masking strategy[21]. In this process, the input node/edge attributes (e.g., atom type in the molecular graph) are randomly masked, and the graph neural network (GNN) is tasked with predicting them. This strategy encourages the model to learn comprehensive representations, which would be instrumental in predicting masked attributes accurately. Therefore, unlike small molecules, designing an appropriate granularity of masking strategies is crucial in peptide representation learning. Because there are various repeated units in the structure of peptides, such as peptide bonds, random masking is not appropriate. For instance, predicting the type of atoms in amino bonds is a simple task, but there are a large number of peptide amino in peptides. Random masking results in the majority of predicted atoms being those in amino bonds. We have designed different masking strategies.

Given that the dataset for non-canonical amino acids is considerably smaller than that of canonical amino acids, we have employed a two-step training approach. The initial phase of pre-training is carried out with data with canonical amino acids, thereby establishing a base for the model to accurately capture the properties of peptide structures. The subsequent phase of pre-training involves data with non-canonical amino acids, which further refines the model, enhancing its ability to recognize the characteristics of a range of novel molecular structures. This method ensures that our model is well-versed in both types of amino acids, thus facilitating efficient feature extraction from peptides that include non-canonical amino acids.

## Heterogeneous Graph-based Peptide Representation

Past research has shown that a heterogeneous graph structure, which combines atoms and fragments, is effective for molecular representation learning[22]. As illustrated in Figure 1, within this heterogeneous graph, atoms and fragments that constitute the peptide are depicted as nodes. Simultaneously, the various relationships—such as the chemical bonds linking atoms, the bonds connecting fragments, and the interactions between atoms and fragments—are represented as edges that connect these nodes. The edges between atoms and fragments represent a subordinate relationship. For instance, a certain fragment only has connections with all the atoms that compose it. The purpose of this design is to allow features of different granularities to be effectively transmitted in message passing mechanism. This method of leveraging a heterogeneous graph structure facilitates the extraction of multi-view peptide molecular graph features, thereby bolstering the representation's capacity to pinpoint and encapsulate essential features and patterns inherent in peptide sequences. Moreover, this approach is conducive to the integration of a plethora of diverse and rich granular information, culminating in the generation of precise and accurate peptide feature representations.

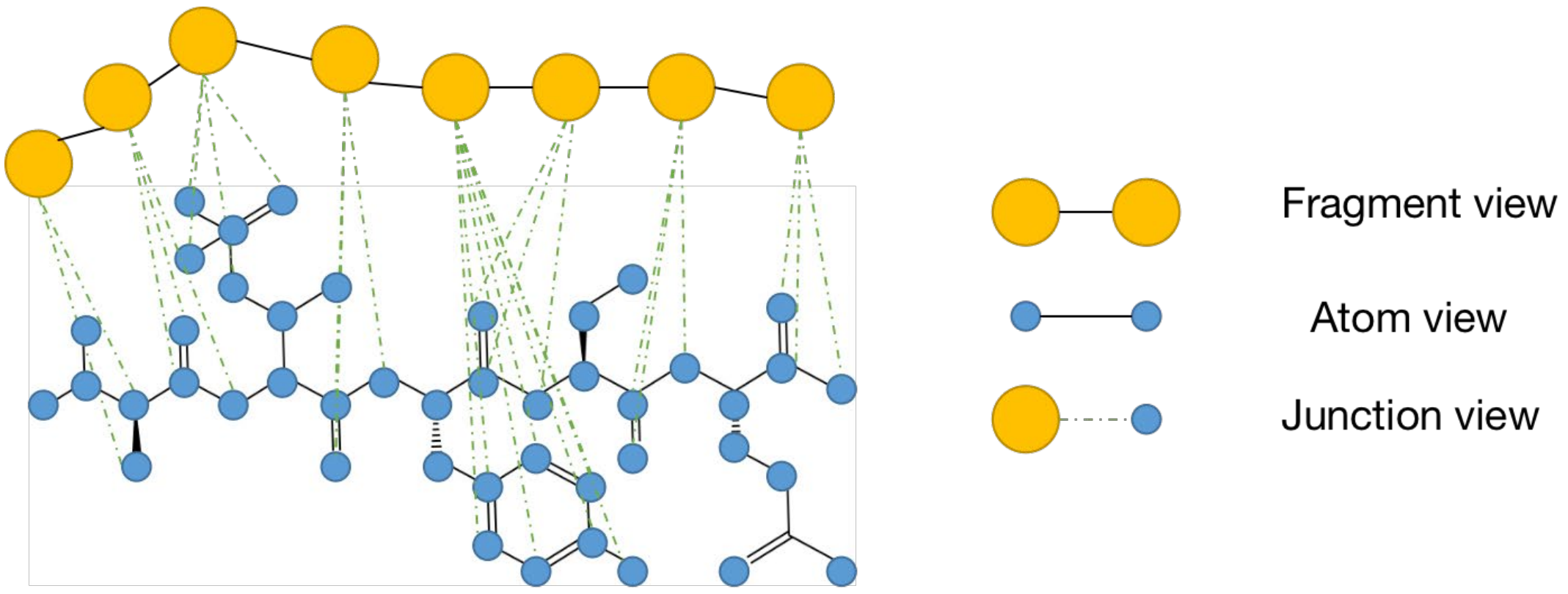

**Figure 1**. The heterogeneous graph of PepLand. Yellow nodes represent fragments, blue nodes represent atoms, and black lines indicate the bonds between nodes of the same type. There's a green dashed line connecting atoms and the fragments they exist in. Therefore, a multi-view heterogeneous graph is constructed, composed of an atomic view, a fragment view, and a cross point view.

In this study, we employ the same message passing mechanism proposed in Pharmacophoric-constrained Heterogeneous Graph Transformer (PharmHGT) [23] to extract features from heterogeneous graphs, shown as Figure2. In comparison to the classical Heterogeneous Graph Transformer (HGT) [24], this method introduces a distinct messaging passing mechanism in graph neural networks (GNNs) that leverages edge features. Specifically, node features are denoted as $X_{V_I}$ , edge features as $X_{e_{ij}}$ , hidden states of edges as $H\left(X_{e_{ij}}\right)$ ,and hidden states of nodes as $H(X_{V_i})$. The messaging steps for information propagation are

as follows:

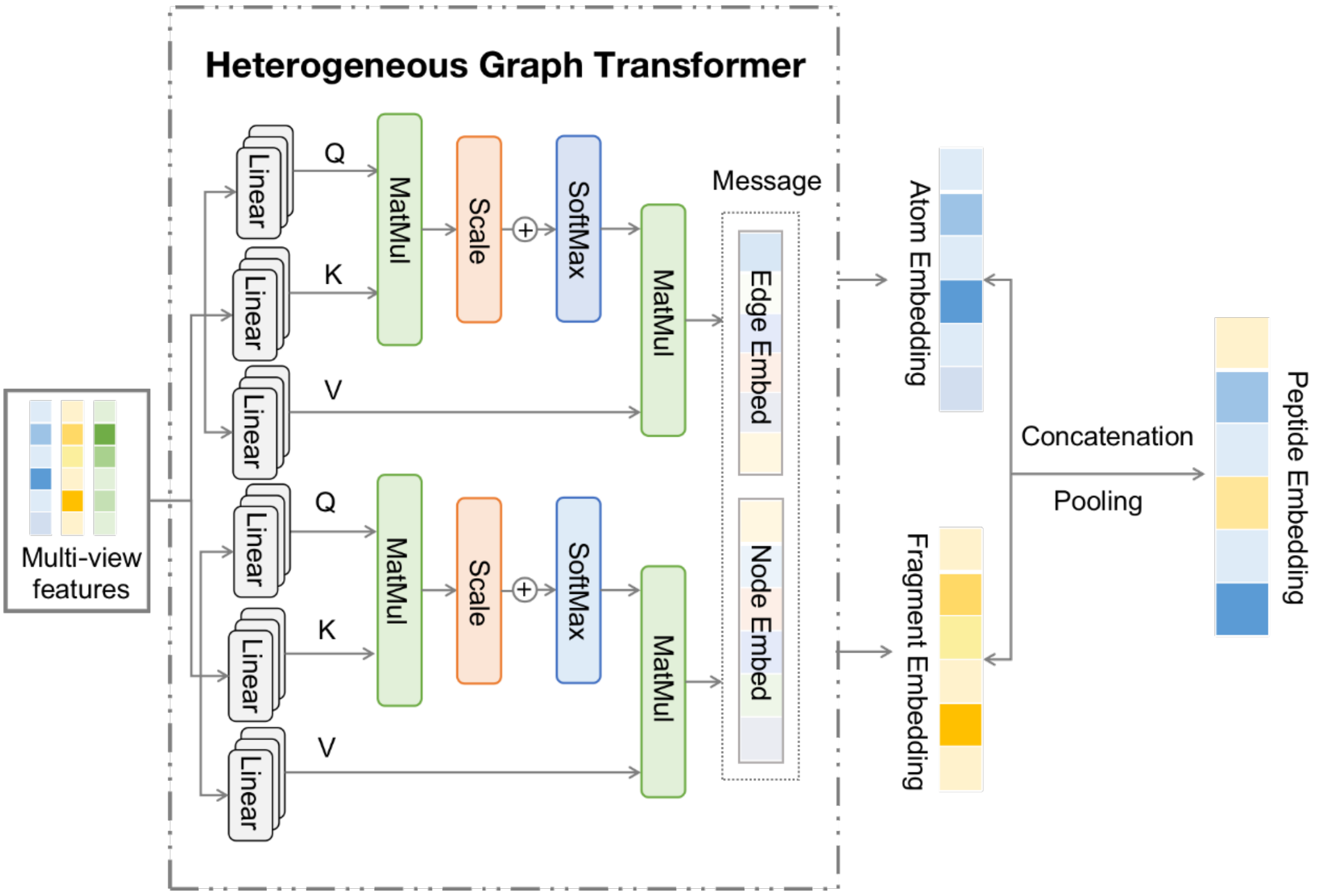

**Figure 2:** The multi-view feature representation learning framework of PepLand

$$M_V^1(X_{vi}) \quad = \quad \sum_{\theta_{\mathcal{N}(V_i)}} H\left(X_{\theta_{\mathcal{N}(V_i)}}\right), \qquad t = 1$$

$$M_E^1\left(X_{e_{ij}}\right) = H\left(X_{v_i}\right), t = 1$$

$$M_V^t(X_{vi}) \quad = \quad \sum_{\theta_{\mathcal{N}(V_i)}} Attention\left(H^{t-1}\left(X_{v_i}\right)\ \right)W_{v_i}^Q, \qquad t > 1$$

$$M^{t-1}\left(X_{\theta_{\mathcal{N}(V_i)}}\right)W_{m_i}^K, H^{t-1}\left(X_{v_i}\right)W_{v_i}^V, \qquad t > 1$$

$$M_E^t\left(X_{e_{ij}}\right) = \quad Linear\left(M_E^1\left(X_{e_{ij}}\right)\right) + \quad H^t(X_{vi}) - H^{t-1}\left(X_{e_{ij}}\right), \qquad t > 1$$

where the $\theta_{\mathcal{N}(V_i)}$ is the function to find edges directed to node $v_i$ and the relevant messages are $M_E^t\left(X_{e_{ij}}\right)$ and $M_V^t(X_{vi})$. Taking into account the problem of vanishing gradients, the model incorporates a basic residual block to stabilize training during multi-view message passing:

$$\begin{cases} H^t(X_{vi}) = H^{t-1}(X_{vi}) + M_V^t(X_{vi}) \\ H^t(X_{e_{vi}}) = H^{t-1}(X_{e_{vi}}) + M_E^t(X_{e_{vi}}) \end{cases}$$

## Fragmentation Method

Previous sections have illustrated that incorporating fragment information and atom-based molecular graphs can enhance the model's representational capacity, particularly for peptide with non-canonical amino acids. However, the definition of a fragment is not unique, they can be large or small. Different types and sizes of fragments play a crucial role in the properties of the final polypeptide [25]. Therefore, we need to meticulously design fragmentation methods suitable for polypeptides which can achieve two objectives. Firstly, our fragments can serve as a supplement and enhancement for existing features that lie between atomic-level and amino acid-level resolution, while incorporating a certain degree of domain knowledge. For example, the six carbons in the benzene ring are almost always considered as a hydrophobic entity [26]. This may benefit graph-based pre-training tasks and various downstream property prediction tasks. Secondly, the fragments library we have should have an appropriate vocab size. The frequency of fragments in the data should not follow a long-tail distribution, as it would make the model's learning more difficult.

In summary, we designed two fragment partitioning strategies:

1. Amiibo(Amino Bond Preservation): Amino bond, a recurring structural unit within the peptide, should be preserved entirely. This yields our first cutting strategy, which is to preserve all the amino bonds as individual fragments. We only break the connection between the amino bond and other functional groups, as depicted in Figure 3.
2. AdaFrag(Adaptive Fragmentation): Amiibo falls short when the side chain of an non-canonical amino acid is very large. In this strategy, side chain groups are further cut using BRICS algorithms [27], promoting information sharing of the same fragment across different side chains and alleviating the problem of long-tail distribution, as depicted in Figure 4.

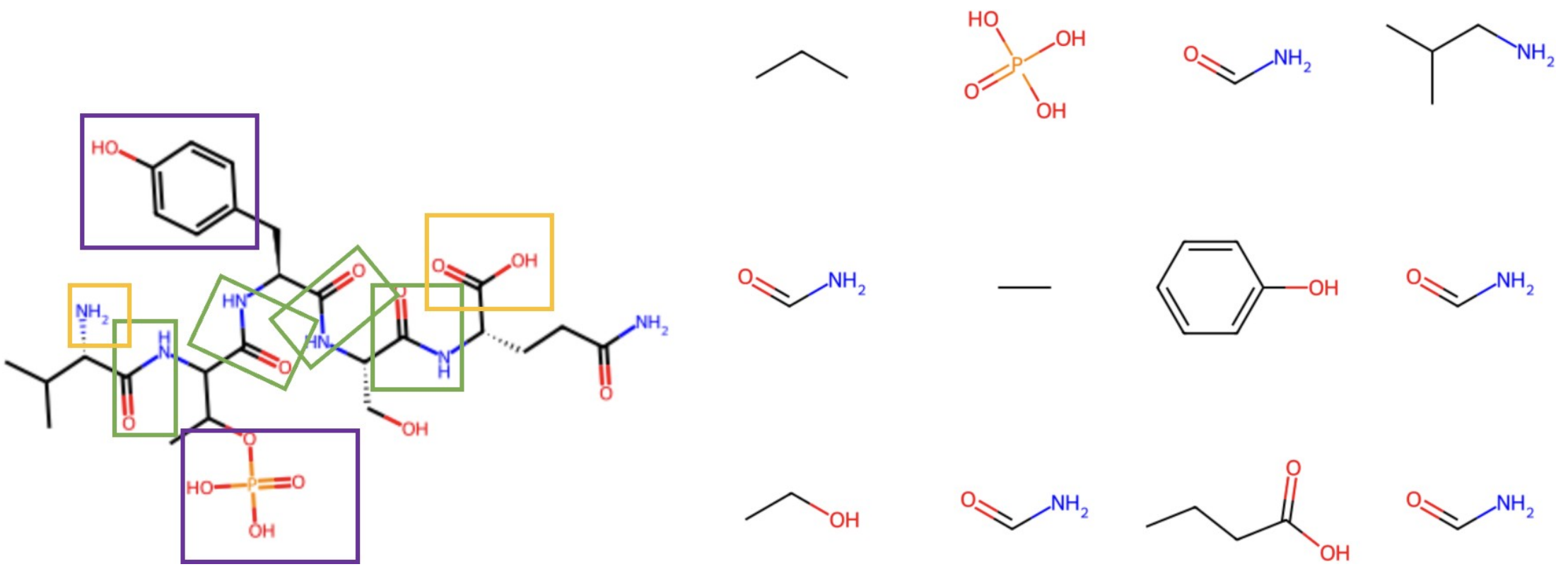

**Figure 3.** Amiibo fragmental strategy: Break the molecules but preserved all amino bonds.

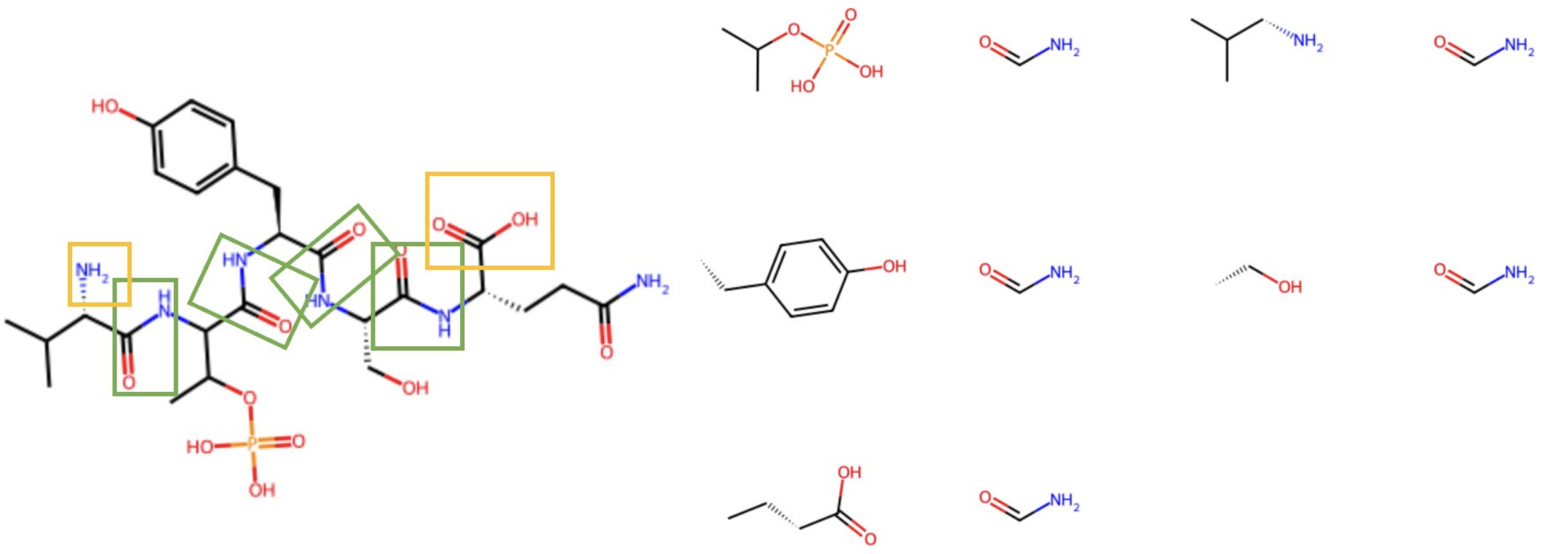

**Figure 4.** AdaFrag fragmental strategy: Break the molecules but preserved all amino bonds and continue breaking the large side chains using BRICS.

Herein, we present the statistical data pertaining to dictionaries derived from two distinct methodologies. In the case of Amiibo strategy, the vocabulary size is comprised of 410 entries, with upwards of 37% of the fragments occurring less than five times, a phenomenon attributable to the relatively coarser granularity employed. Conversely, AdaFrag strategy encompasses a vocabulary of 258 entries, with a mere 6% of the fragments manifesting less than five occurrences. It is important to note that fragments with a frequency of two or fewer have been expunged from the dictionaries. Moreover, the recovery rate serves as a crucial metric in the assessment of a fragment method. In this context, the recovery rate is defined as the efficacy of reconstituting the fragments into the original molecule, as facilitated by a specific fragment method. Notably, Amiibo strategy exhibits a diminished recovery rate of 88.6%, a consequence of the fragment library conforming to a long-tailed distribution in frequency, which results in the exclusion of many low-frequency fragments situated in the distribution's tail.

**Table 1**: the statistical data for the fragment library derived from two fragmentation strategies.

| | Vocab Size | Ratio of Frequency <= 5 | Recovery Rate |
|---|---|---|---|
| Amiibo | 410 | 37.10% | 88.60% |
| AdaFrag | 258 | 6.10% | 97.52% |

## Masking Method

Inspired by previous research [21], we devised several self-supervised learning pre-training strategies. Similar to masked language modeling (MLM) that masks out some tokens from the input sentences and then trains the model to predict the masked tokens by the rest of the tokens, the proposed pre-training strategy in this study involves masking certain nodes from the heterogeneous graph structure, including atoms and fragments, and then training the model to predict them [28]. However, because the repeated units and specific substructures in the polypeptide are related to specific properties, random masking which is commonly used in MLM is not a suitable strategy. For example, amino acids are basically formed by an amino group, a carboxyl group, and a side chain. Random masking could potentially mask a portion of the amino or carboxyl group, and the model would predict these atoms quite easily.Therefore, we designed the following masking strategies and compare their performance:

1. RandomMasking: This strategy involves randomly masking individual atoms within the molecular graph, forcing the model to learn representations that can predict the missing atom identities.
2. BulkMasking: BulkMasking strategy, which involves masking all the atoms constituting a single amino acid. This approach reduces the availability of neighboring atom information and increases the difficulty of prediction. This method is a bit like the practice of masking amino acids in protein language model training. The only difference is that we are masking not only canonical amino acids, but also non-canonical amino acids.
3. SideChainMasking: Since atoms in peptide bonds constitute a significant portion of protein sequences, Random Atom Masking and Bulk Atom Masking has a high probability of selecting atoms within peptide bonds. This significantly reduces the difficulty of the pre-training task. Therefore, this strategy further selectively masks atoms in the side chains while preserving the atoms in the peptide bonds.

4. FragmentMasking: In this strategy, we mask entire fragments within the peptide molecule, encouraging the model to capture the relationships and properties associated with the masked fragments.

## Two-step Pre-training Strategy

In this study, we present a dual-phase training approach for the representation learning of peptides containing non-canonical amino acids. Initially, the model undergoes training using a dataset exclusively composed of canonical amino acids until it achieves convergence. These amino acid sequences, approximately 8 million in number, have been meticulously extracted from the Uniprot Database. We have classified protein sequences containing fewer than 50 amino acids as peptide sequences. This preliminary phase facilitates the model in assimilating evolutionary data from an extensive array of peptide sequences, a process analogous to that employed by protein language models. However, considering the disproportionate ratio of proteins with non-canonical amino acids to the plethora of non-canonical amino acids, a second training phase is warranted. This phase is dedicated to isolating and assimilating the distinctive attributes of non-canonical amino acids. Here, the model is trained solely on datasets comprised of peptides with non-canonical amino acids, with the objective of distilling the intricate information encapsulated within these protein sequences.

## Implementation Details

All the experiments were conducted on a Linux server with 4 CPUs (20 cores per CPU), 4 GPU cards (Nvidia 3080, 24 GB memory per card), and 256 GB of system memory. The source code was implemented using the programming language Python version 3.10.0 with the package PyTorch version 1.10.1 and torch-geometric 2.0.3.

# Experimental Results

## Experimental setup

For the proposed pre-training method in this paper, we set the batch size to 512. Due to the simplicity of the masked task and the fast convergence of the model, training was conducted for 1 epoch with a learning rate of 0.001. The embedding sizes for atoms and fragments were set to 300, and the graph transformer had 5 layers. To ensure fairness, all downstream tasks in the experiments were conducted with the same experimental settings, as follows: the affinity task required 200 epochs for convergence, while the remaining downstream tasks required 100 epochs. The learning rate for the CPP Permeability task was set to 0.00001, and 0.00005 for the other tasks. Additionally, the batch size for all tasks was set to 32.

To our knowledge, there is currently no dedicated training model specifically for peptides, especially involving non-canonical amino acids. In order to compare with tasks related to the prediction of non-canonical peptide properties, we utilized models from the small molecule domain as our comparison baselines. To make the comparison fairer, we selected three different types of pre-trained models that utilize 1D, 2D, and 3D molecular information, respectively. These models include Chemberta2[29], a sequence-based language model trained on one of the largest small molecule datasets; MolCLR[30], a pre-trained graph neural network that employs a molecular contrastive learning strategy; and Uni-Mol[31], a universal 3D MRL framework. For tasks involving only canonical amino acids, we used previous models as well as protein language models, such as ESM2 [29] and ProteinBert[9], as comparison baselines. We used the largest version of ESM2 which has 48 transformer layers.

We evaluate our proposed pretraining methods PepLand with the baseline models on all evaluating datasets. The classification task was assessed using ROC-AUC, F1, and Accuracy metrics, whereas the regression task predictions was gauged using the Pearson correlation coefficient and Spearman rank correlation coefficient.

## Pooling method evaluation

The proposed PepLand is a pre-training method that employs a graph neural network (GNN) model. This model generates embeddings at the atomic and fragment levels instead of providing a direct representation of protein features. To obtain a comprehensive peptide representation with rich informative content, we explored various graph pooling approaches. Specifically, we evaluated the performance of three distinct graph pooling methods: max pooling, average pooling, and GRU pooling. These methods were assessed based on their ability to capture essential information and generate meaningful protein representations. As shown in Fig.6, the outcomes of the experiments revealed that the GRU pooling method surpassed the others in all downstream tasks. This was especially evident in the Non-canonical CPP task, where the Spearman correlation reached 0.742, and the Pearson correlation reached 0.670. For further details, please refer to the supplementary materials.

To ensure that the experimental results accurately reflect the capability of the trained model itself, we only connected two simple linear layers to the representations obtained from all pre-trained models for downstream tasks. This approach was chosen to eliminate any potential confounding variables that could arise from more complex downstream architectures, thus providing a more reliable assessment of the pre-trained model's performance.

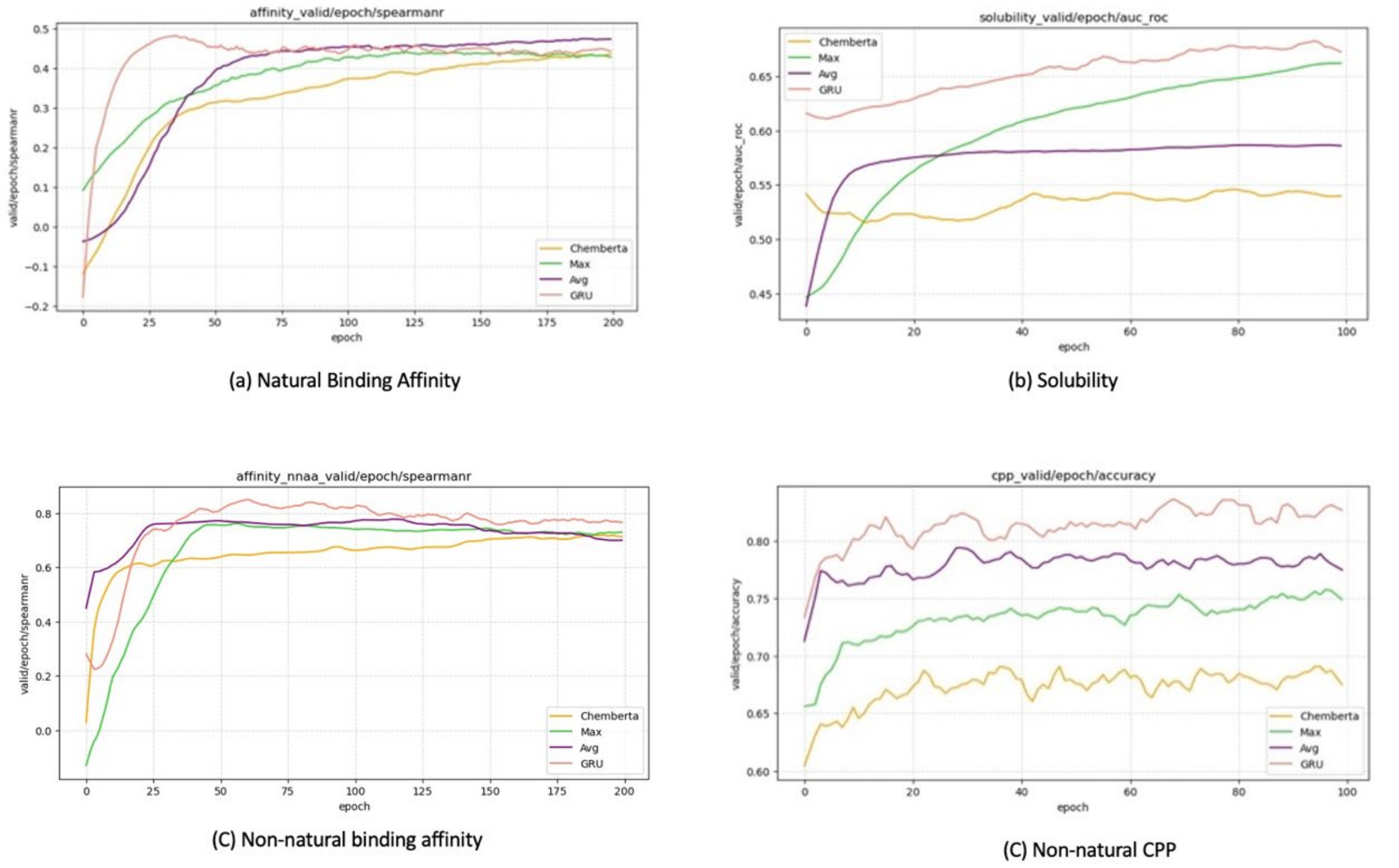

**Figure 5.** The selection of various graph pooling methods has a substantial impact on downstream tasks. The figure showcases the effects of four pooling techniques, namely max pooling, average pooling, and GRU, on five distinct downstream tasks.

## Performance comparison

As can be seen in Fig.5, our model exhibited considerable enhancement in performance compared to the baseline models in two tasks related to non-canonical amino acids. In the non-canonical binding affinity prediction task, our proposed model achieved a Spearman correlation of 0.768 and a Pearson correlation of 0.873, while Chemberta2 obtained a Spearman correlation of 0.721 and a Pearson correlation of 0.792. In comparison, PepLand demonstrated improvements of 6.516% and 10.227% over Chemberta2 in these two metrics, respectively. It is worth noting that Uni-Mol and MolCLR underperformed in the peptide binding affinity prediction task, potentially due to their reliance on extensive structural information from small molecules, which differs in data distribution from that of peptides.

In the non-canonical cell penetrating prediction task, PepLand achieved a Spearman correlation of 0.628 and a Pearson correlation of 0.520, marking an improvement of 18.45% and 16.51%, respectively, compared to the second-best baseline, Uni-Mol. This clearly indicates that PepLand significantly surpasses other baselines in predicting cell penetration tasks, thereby highlighting its superior ability to extract feature representations of non-canonical amino acids, which pose a challenge for other models to capture effectively.

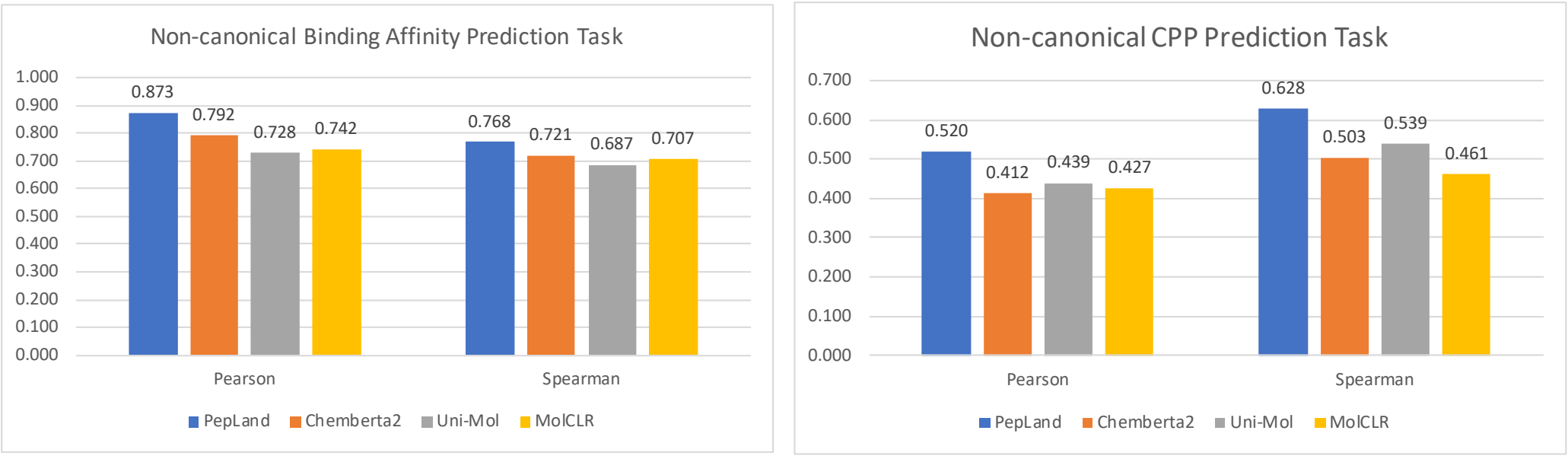

**Figure 6.** Performance comparison on Non-canonical binding affinity prediction task and Non-canonical cell penetrating prediction task.

**Table 2.** Performance comparison between peptide task with only canonical amino acid. Better results are indicated in bold.

| | Canonical CPP | | | Solubility | | | Canonical Binding Affinity | | | |
|---|---|---|---|---|---|---|---|---|---|---|
| | Acc | F1 | AUC | Acc | F1 | AUC | MAE | RMSE | PCC | SCC |
| Uni-Mol | 0.704 | 0.718 | 0.732 | 0.604 | 0.657 | 0.646 | 1.165 | 1.258 | 0.432 | 0.411 |
| MolCLR | 0.691 | 0.714 | 0.728 | 0.578 | 0.597 | 0.632 | 1.171 | 1.475 | 0.424 | 0.446 |
| Chemberta2 | 0.711 | 0.747 | 0.773 | 0.563 | 0.693 | 0.549 | 1.183 | 1.508 | 0.437 | 0.488 |
| ESM | 0.815 | 0.823 | 0.885 | 0.662 | 0.708 | 0.725 | **0.958** | **1.236** | 0.446 | 0.452 |
| PepLand | 0.767 | 0.772 | 0.838 | 0.616 | 0.69 | 0.662 | 1.146 | 1.493 | 0.45 | **0.503** |
| PepLand + ESM | **0.828** | **0.839** | **0.885** | **0.669** | **0.712** | **0.73** | 1.123 | 1.461 | **0.53** | 0.454 |

In tasks related to canonical amino acids alone (CPP and Solubility), our model surpassed other small molecule models such as Uni-Mol, MolCLR, and Chemberta2 in all evaluated metrics, except it fell short when compared to the protein language model ESM2. This might be due to ESM2 being trained on the UniRef50 dataset, which comprises 45 million protein sequences, making its parameter count reach 15 billion, significantly larger than that of PepLand. PepLand was trained only on a dataset of 8 million peptides, thus it did not capture evolutionary information on a large scale. However, ESM is incapable of handling non-canonical amino acids, a capability that our PepLand model holds, as ESM's input is designed for standard amino acid sequences. Given PepLand prowess in handling peptides with non-standard amino acids, we attempted to merge the strengths of both ESM and PepLand in tasks concerning canonical amino acids. This resulted in notable enhancements, achieving an AUC of 0.885 in the CPP task and an AUC of 0.730 in the Solubility task. In the Affinity task, our model realized a Spearman correlation of 0.454 and a Pearson correlation of 0.530.

## The Importance of granularity

To investigate the true impact of granularity, a detailed comparative experiment was conducted. Two other fragmentation methods were employed for comparison with PepLand. Specifically, Molecular Graph was used, representing an atomic granularity akin to not employing any fragmentation method. The other method, Principal Subgraph [32], is a data-driven fragmentation method capable of automatically discovering frequent principal subgraphs from the dataset. To ensure a fair comparison, the same training and testing procedures were maintained, only altering the graph construction process. For Molecular Graph, since there's only one type of node representing atoms, there is only a single view.

Overall, employing the atomic granularity of Molecular Graph yielded the poorest results. This might be due to the larger molecular graph structures of polypeptides, leading to information loss during the message passing and pooling processes. Principal subgraph, despite being a fragmentation method without integrated domain knowledge, exhibited decent performance. However, due to the lack of relevant prior knowledge, its modeling capability for polypeptides containing non-canonical amino acids was still be inferior to PepLand.

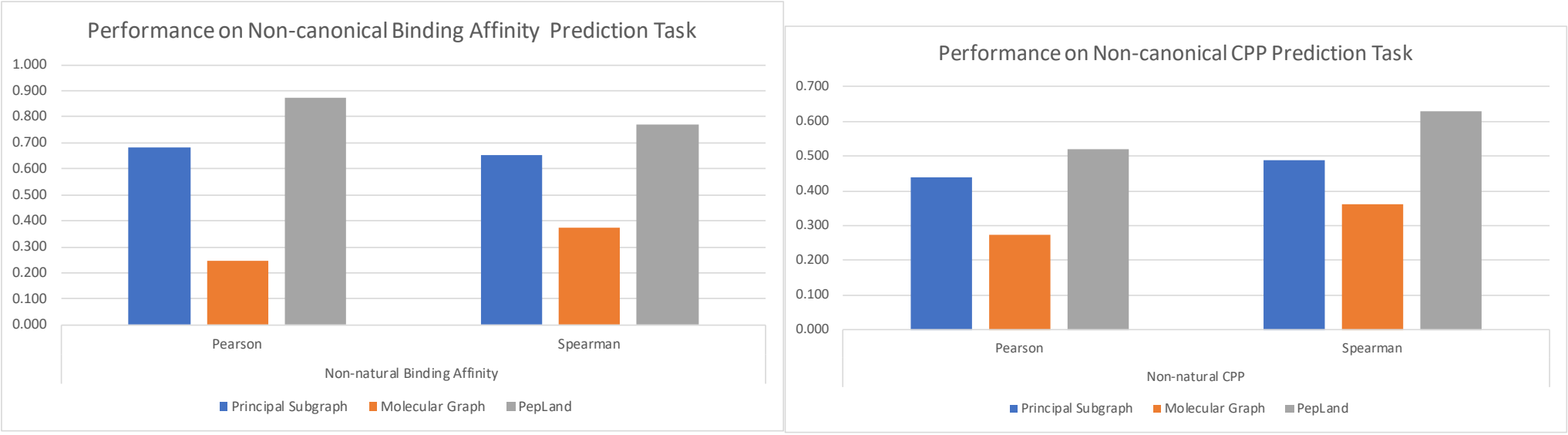

**Figure 7.** The performance on two prediction tasks of different fragmentation methods.

## Exploring impact on fragmental strategies

The impact of the fragmental strategy on downstream tasks was studied. The Tabel3 demonstrates the performance metrics of different strategies across five down- stream tasks. Amiibo and AdaFrag represent the two fragmental strategies proposed in previous section.

From the experimental results, it can be seen that AdaFrag performs better on most tasks. As shown in Table 3, although the F1 score of AdaFrag in Solubility prediction task is slightly lower than Amiibo, it is 14.39% higher than Amiibo in terms of ROC-AUC metric. In tasks involving non- amino acids, AdaFrag's Spearman correlation and Pearson correlation are both much better than Amiibo. Considering the analysis of the dictionaries obtained from the two strategies in the previous chapters and the results of the ablation experiments, we believe that AdaFrag is better at breaking peptides into appropriate fragments.

**Table 3.** Experiments for the impact of the fragmental strategy on downstream tasks. Amiibo means to keep the amino bond and cut the connection between the amino bond and other functional groups. AdaFrag means to use the BRICS algorithm to further cut the side chain based on Amiibo. Better results are indicated in bold.

| Dataset | Metrics | Amiibo | AdaFrag |
|---|---|---|---|
| Canonical CPP | Acc | 0.746 | 0.763 |
| | F1 | 0.752 | 0.78 |
| | AUC | 0.787 | 0.82 |
| Solubility | Acc | 0.556 | 0.616 |
| | F1 | **0.696** | **0.693** |
| | AUC | 0.556 | 0.636 |
| Canonical Affinity | MAE | 1.169 | 1.15 |
| | RMSE | 1.521 | 1.5 |
| | PCC | 0.476 | 0.496 |
| | SCC | 0.44 | 0.461 |
| Non-canonical Affinity | MAE | **1.064** | **1.095** |
| | RMSE | **1.4** | **1.437** |
| | PCC | 0.807 | 0.862 |
| | SCC | 0.708 | 0.749 |
| Non-canonical CPP | MAE | **0.649** | **0.622** |
| | RMSE | 1.031 | 1.003 |
| | PCC | 0.426 | 0.477 |
| | SCC | 0.506 | 0.556 |

## Exploring impact on masking strategies

We investigated the impact of the masking methods on downstream tasks. Here are four masking strategies which are RandomMasking, BulkMasking, and SideChainMasking and FragmentMasking as detailed in previous section . The results are shown in the Table4.

Masking only the atoms on the side chains increases the task difficulty, allowing the model to learn more evolutionary information. The results in the Table 3 demonstrate that the method of random fragment masking strategy performs well in different downstream tasks, particularly in tasks involving non-canonical amino acids. The Spearman correlations for the affinity and Non-nocanonical CPP tasks with non-canonical amino acids are 0.767 and 0.589, respectively.

**Table 4.** Experiments for the impact of the masking methods on downstream tasks. Better results are indicated in bold.

| Dataset | Metrics | RandomMasking | BulkMasking | SideChainMasking | FragmentMasking |
|---|---|---|---|---|---|
| CPP | Acc | 0.733 | 0.763 | **0.784** | 0.767 |
| | F1 | 0.763 | 0.78 | **0.786** | 0.772 |
| | AUC | 0.812 | 0.82 | 0.833 | **0.838** |
| Solubility | Acc | 0.563 | 0.616 | 0.57 | **0.589** |
| | F1 | **0.7** | 0.693 | 0.69 | 0.691 |
| | AUC | 0.576 | **0.636** | 0.601 | 0.6 |
| Canonical Binding Affinity | MAE | 1.155 | 1.15 | 1.15 | **1.133** |
| | RMSE | **1.496** | 1.5 | 1.547 | 1.504 |
| | PCC | 0.496 | 0.496 | 0.445 | 0.5 |
| | SCC | 0.452 | 0.461 | 0.459 | **0.481** |
| Non-canonical Binding Affinity | MAE | 1.039 | 1.095 | 0.919 | **0.9** |
| | RMSE | 1.486 | 1.437 | 1.228 | **1.21** |
| | PCC | 0.756 | **0.862** | 0.83 | 0.827 |
| | SCC | 0.678 | 0.749 | 0.708 | **0.767** |
| Non-canonical CPP | MAE | 0.615 | 0.622 | **0.641** | 0.61 |
| | RMSE | **0.993** | 1.003 | 1.018 | 0.996 |
| | PCC | **0.555** | 0.477 | 0.448 | 0.483 |
| | SCC | 0.506 | 0.556 | 0.518 | **0.589** |

## Exploring impact on training strategies

The effect of the two-step pre-training on downstream tasks was tested. We performed both the two-step pre training and single step pre-training and results on the five downstream tasks indicate that the two-step pre-training facilitates the extraction of richer representation information of

In Table 4, it can be seen that the two-step training strategy achieves better results on all benchmarks. Two-step pre-training strategy specifically designed for non-canonical amino acids can learn more targeted representations, thereby improving the prediction accuracy of downstream tasks. This improvement is particularly evident in tasks involving non- canonical amino acids, with Pearson correlations of 0.873 and 0.520 achieved for the non- canonical binding affinity and Non- canonical CPP tasks, respectively.

**Table 5.** Experiments for the effect of the single-step and two-step pre-training strategy on down-stream tasks.

| Dataset | Metrics | One-step | Two-step |
|---|---|---|---|
| CPP | Acc | 0.767 | **0.767** |
| | F1 | **0.777** | 0.772 |
| | AUC | 0.836 | **0.838** |
| Solubility | Acc | 0.589 | **0.616** |
| | F1 | **0.691** | 0.69 |
| | AUC | 0.6 | **0.662** |
| Canonical Binding Affinity | MAE | 1.155 | **1.146** |
| | RMSE | 1.504 | **1.493** |
| | PCC | **0.496** | 0.45 |
| | SCC | 0.481 | **0.503** |
| Non-Canonical Binding Affinity | MAE | 0.9 | **0.762** |
| | RMSE | 1.21 | **1.045** |
| | PCC | 0.827 | **0.873** |
| | SCC | 0.767 | **0.768** |
| Non-Canonical CPP | MAE | 0.61 | **0.59** |
| | RMSE | 0.996 | **0.97** |
| | PCC | 0.483 | **0.52** |
| | SCC | 0.589 | **0.628** |

## Exploring impact on multi-view features

The impact of multi-view feature selection on downstream tasks was studied. For the three views included in the model, we explored various combinations of multi-view features to identify the most reasonable feature representation. Considering the performance across the five downstream tasks, the combination of atom embedding with atom-view and junction-view information proves to be the most effective, while the fragment embedding only with fragment-view information yields the best results.

Regarding the selection of multi-view information, the embeddings of atom nodes combine atom-level and junction-level information, while the embeddings of fragment nodes only include fragment-level information. This view selection demonstrates good performance in different downstream tasks, as shown in the Table 5. In the affinity task with non-canonical amino acids, the Pearson correlation reaches 0.876.

**Table 6.** Experiments for the impact of multi-view feature selection on downstream tasks

| Dataset | Metrics | A & F | AJ & F | AJ & FJ |
|---|---|---|---|---|
| CPP | Acc | **0.789** | 0.767 | 0.759 |
| | F1 | **0.809** | 0.777 | 0.769 |
| | AUC | **0.844** | 0.836 | 0.819 |
| Solubility | Acc | 0.582 | 0.589 | **0.589** |
| | F1 | 0.690 | **0.691** | 0.690 |
| | AUC | 0.602 | 0.600 | **0.628** |
| Canonical Binding Affinity | MAE | 1.157 | **1.155** | 1.177 |
| | RMSE | 1.526 | **1.504** | 1.526 |
| | PCC | 0.470 | **0.496** | 0.466 |
| | SCC | 0.433 | **0.481** | 0.442 |
| Non-Canonical Binding Affinity | MAE | 1.322 | 0.900 | **0.817** |
| | RMSE | 1.728 | 1.210 | **1.055** |
| | PCC | 0.651 | 0.827 | **0.876** |
| | SCC | 0.695 | 0.767 | **0.769** |
| Non-Canonical CPP | MAE | 0.604 | **0.610** | 0.608 |
| | RMSE | **0.977** | 0.996 | 0.980 |
| | PCC | **0.524** | 0.483 | 0.511 |
| | SCC | 0.581 | 0.589 | **0.591** |

For more detaails, please refer to the supplementary materials.

## Case Study1: Binding Affinity Prediction of Cyclic Peptides

A case study was conducted on cyclic peptides, which are polypeptides containing non-canonical amino acids, due to their promising pharmacological potential. We utilized

Haddock [33], Rosetta FlexDDG [34], and DeepPurpose [35] as baselines for affinity prediction, all of which are open-source methods. HADDOCK specializes in the flexible docking for biomolecular complex modeling, while Rosetta FlexDDG, part of the Rosetta macromolecular modeling suite, samples conformational diversity to estimate interface $\Delta\Delta G$ values. DeepPurpose is a flexible deep learning-based framework for predicting protein-ligand interactions.

To our knowledge, there aren't established benchmarks for cyclic peptide affinity prediction. We curated a set of cyclic peptide designs associated with two proteins from SKEMPI 2.0 database [36]. The proteins, identified by PDB IDs 5XCO and 1SMF, are related to the crystal structure of human K-Ras G12D mutant with GDP and cyclic inhibitory peptides, and artificial trypsin inhibitor peptides, respectively. A total of 10 peptides from K-ras G12D Mutant and 6 from 1SMF were included, with corresponding $\Delta\Delta G$. To prevent data leakage, we excluded these data from our training dataset, and post-training, tested on these two independent sets. Haddock and Rosetta FlexDDG required no training, whereas DeepPurpose was trained with the mentioned dataset.

Notably, excellent prediction results were achieved on 1SMF with a Spearman Correlation of 0.829, indicating strong true binder recall capability, crucial for virtual screening in drug discovery process. Rosetta FlexDDG also performed well on 1SMF with a Spearman correlation of 0.77, but Haddock and DeepPurpose lagged with correlations of 0.19 and 0.26, respectively. On the 5XCO dataset, we attained a Spearman Correlation of 0.718 and in our predicted top 3 candidates, two of them are actually ranked within the top 3 based on experimental DDG. However, both Haddock and DeepPurpose performed poorly, scoring only 0.19 and 0.26 in Spearman Correlation, while Rosetta could not compute due to the deletion of amino acid in mutants in this dataset. Overall, PepLand significantly outperformed current CADD(Computer-Aided Drug Design) and other deep learning-based methods in predicting cyclic peptide affinity, showcasing its superior accuracy and reliability.

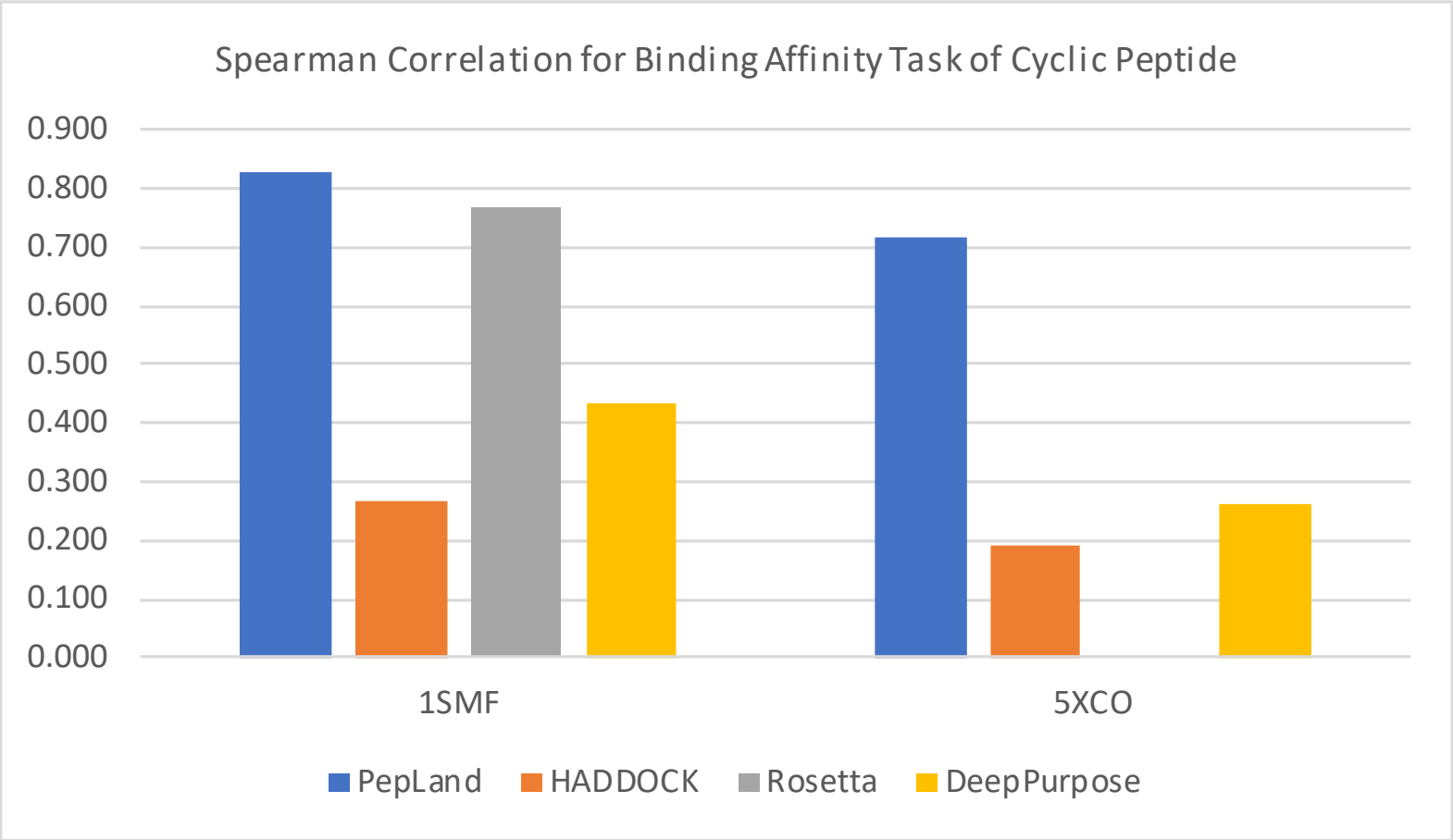

**Figure 8.** Spearman correlation achieved by PepLand, HADDOCK, Rosetta and DeepPurpose on binding affinity prediction task of cyclic peptides.

## Case Study2: Peptide Synthesizability Prediction

In the process of chemically synthesizing polypeptides, amide bonds are formed step-by-step on a solid support that's immobilized. Achieving high yields is crucial for the effective addition of each amino acid to the growing chain, but this can be hindered by sequence-dependent events like aggregation. [37] proposes using a deep learning approach, where ECFP fingerprints are used as input on a dataset collected from an Automated Fast-Flow Peptide Synthesis (AFPS) setup to predict the synthesis of peptides and sequence-specific events. These events are marked by issues in mass transfer and slow reaction kinetics, showing up as broader, flattened UV-vis deprotection peaks.

The output of the model includes Area, Height, and Width, with a particular metric mentioned in the paper being the difference of normalized width minus normalized height (W − H), to quantify such events. Since X didn't provide a pre-trained model, a new model was trained based on the code provided by [37], and tested on it's benchmark. Here, features extracted by PepLand were compared with ECFP fingerprints. The features from PepLand reduced the Mean Absolute Error (MAE) of predicting the difference in peak height and width from 0.160 to 0.138 and improved the coefficient of determination(R2) from 0.644 to 0.746, increased by 15.84%. This indicates that the pre-trained PepLand model is superior in extracting features, capturing structural information of amino acids and polypeptide sequences more accurately, and improving the precision in predicting the difference in peak height and width during the deprotection process. This showcases the potential application of the PepLand model in the field of polypeptide chemical synthesis.

**Table 7.** The performance comparison between using ECFP fingerprints and PepLand.

| | MAE | | | | R2 | | | |
|---|---|---|---|---|---|---|---|---|
| | Area | Hight | Width | H-W | Area | Hight | Width | H-W |
| ECFP | 0.12 | 0.14 | 0.1 | 0.16 | 0.543 | 0.51 | 0.637 | 0.644 |
| PepLand | **0.095** | **0.103** | **0.063** | **0.138** | **0.649** | **0.698** | **0.723** | **0.746** |

# Conclusion

In conclusion, our work pioneers the first pre-trained model, PepLand, in the realm of peptide synthesis, boasting the ability to process not only canonical but also non-canonical amino acids, marking a significant stride in this domain. The innovative fragment method we devised for polypeptides, as evidenced by our experiments, significantly aids in dissecting the granularity of polypeptides, thereby facilitating the model in learning the properties of polypeptides effectively. Our utilization of a multi-view heterogenous graph network alongside a two-step learning strategy underscores the robustness of our approach. Additionally, the introduction of a Masking training method tailored for peptide data significantly contributes to our model's performance. The extensive ablation studies conducted affirm the rationality behind our model's structure, demonstrating that each aspect of PepLand is instrumental in enhancing its overall performance.

Furthermore, our model has proven its mettle across numerous downstream tasks through a myriad of experiments. It shines exceptionally in tasks like affinity prediction, solubility prediction, and transmembrane prediction, particularly excelling in predicting the properties of synthetic peptides containing non-canonical amino acids, where PepLand distinctly leads.

Despite these strengths, we acknowledge certain limitations in our work. Firstly, the training data utilized may not encompass the entire diversity of synthetic peptides, especially those with rare or newly discovered non-canonical amino acids. Although the fragmentation method exists and can handle some of the non-natural amino acids not covered by known data, it cannot fully address this issue. Secondly, our training set comprises less than 10,000 synthetic peptides, with a majority being transmembrane peptides. We anticipate that the accumulation of more data in this domain will further refine and enhance the performance and applicability of PepLand.

The case studies we conducted, especially the promising results in predicting cyclic peptides and the synthesizability of polypeptides, underscore the effectiveness and the broad spectrum of applications that PepLand can cater to. With the successful development and validation of PepLand, we have laid down a solid foundation for leveraging machine learning in peptide research involving non-canonical amino acids. This venture, we believe, will catalyze the advent of more sophisticated models and applications, propelling the exciting frontier of peptide representation further into uncharted territories.

## Acknowledgements

This publication is based upon work supported by the King Abdullah University of Science and Technology (KAUST) Office of Research Administration (ORA) under Award No URF/1/4352-01-01, FCC/1/1976-44-01, FCC/1/1976-45-01, REI/1/5234-01-01, REI/1/5414-01-01, REI/1/5289-01-01, REI/1/5404-01-01. This work is also supported by the Senior and Junior Technological Innovation Team (20210509055RQ), the Jilin Provincial Key Laboratory of Big Data Intelligent Computing (20180622002JC), and the Fundamental Research Funds for the Central Universities, JLU.

# PepLand: a large-scale pre-trained peptide representation model for a comprehensive landscape of both canonical and non-canonical amino acids

Ruochi Zhang[1,2,3], Haoran Wu[3], Yuting Xiu[3], Kewei Li[1,4], Ningning Chen[3], Yu Wang[3], Yan Wang[1,2,4], Xin Gao[5,6,*], Fengfeng Zhou[1,4,*].

1 Key Laboratory of Symbolic Computation and Knowledge Engineering of Ministry of Education, Jilin University, Changchun, Jilin, China, 130012.

2 School of Artificial Intelligence, Jilin University, Changchun, China,130012.

3 Syneron Technology, Guangzhou, China, 510700.

4 College of Computer Science and Technology, Jilin University, Changchun, Jilin, China, 130012.

5 Computational Bioscience Research Center, King Abdullah University of Science and Technology (KAUST), Thuwal, Saudi Arabia, 23955.

6 Computer Science Program, Computer, Electrical and Mathematical Sciences and Engineering Division, King Abdullah University of Science and Technology (KAUST), Thuwal, Saudi Arabia, 23955.

# Correspondence may be addressed to Fengfeng Zhou: FengfengZhou@gmail.com or ffzhou@jlu.edu.cn . Lab web site: http://www.healthinformaticslab.org/. Phone: +86-431-8516-6024. Fax: +86-431-8516-6024. Correspondence may also be addressed to Xin Gao: xin.gao@kaust.edu.sa.

# Supplementary Note 1: Ablation Study

This section serves as a supplementary figure for the experimental results of the ablative experiments.

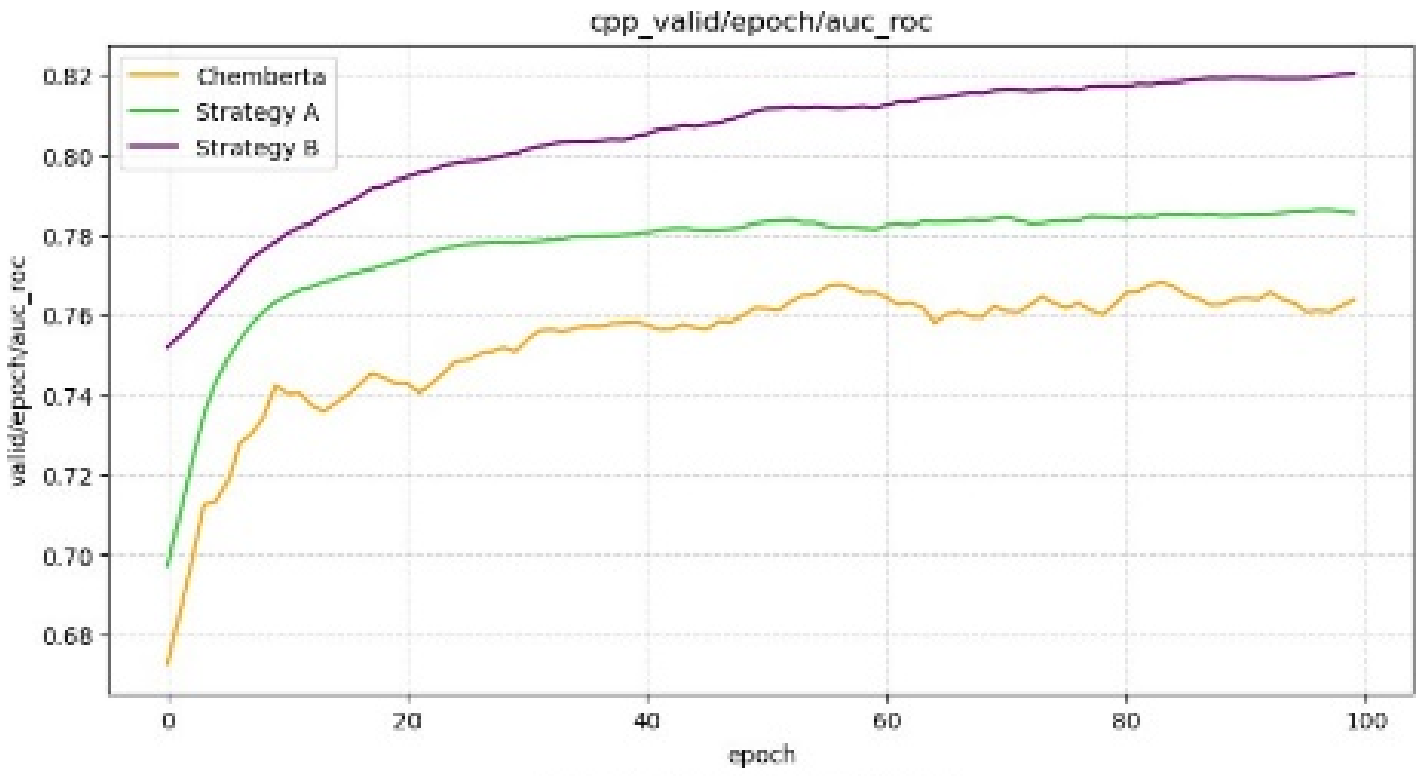

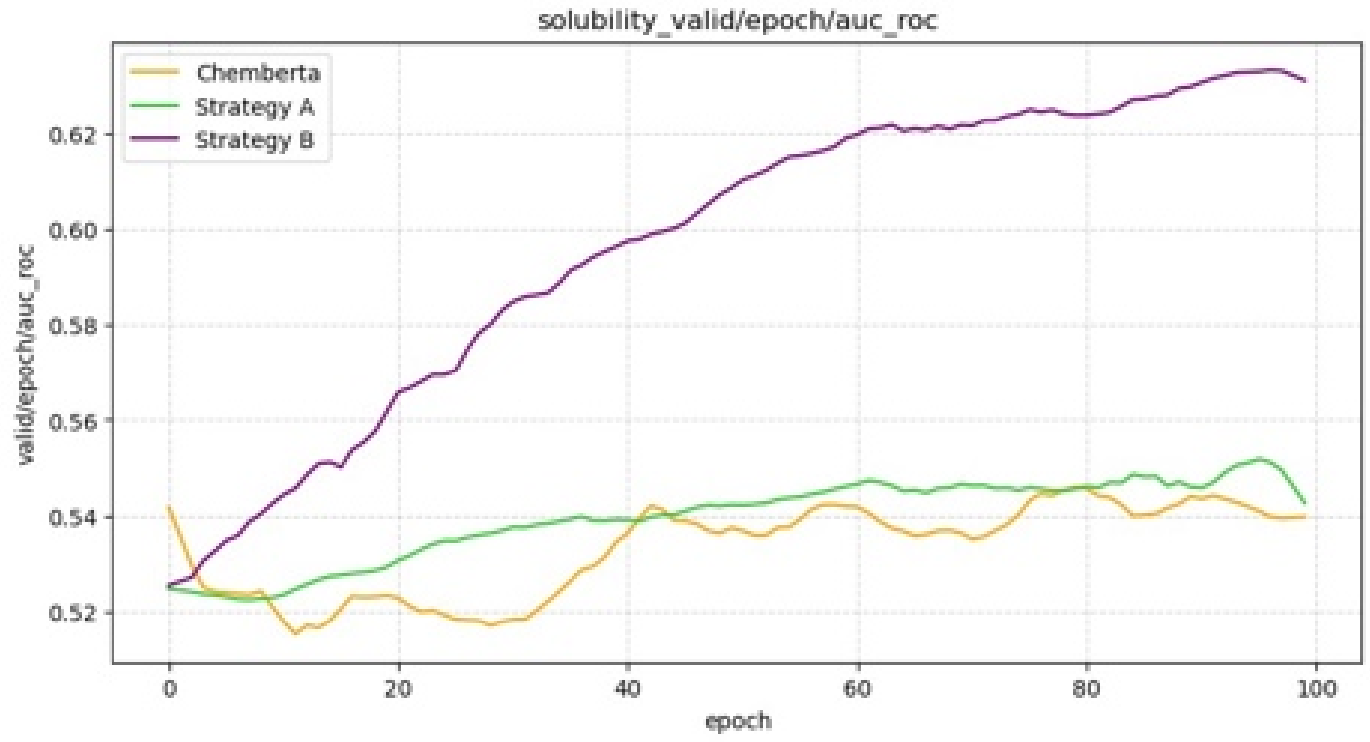

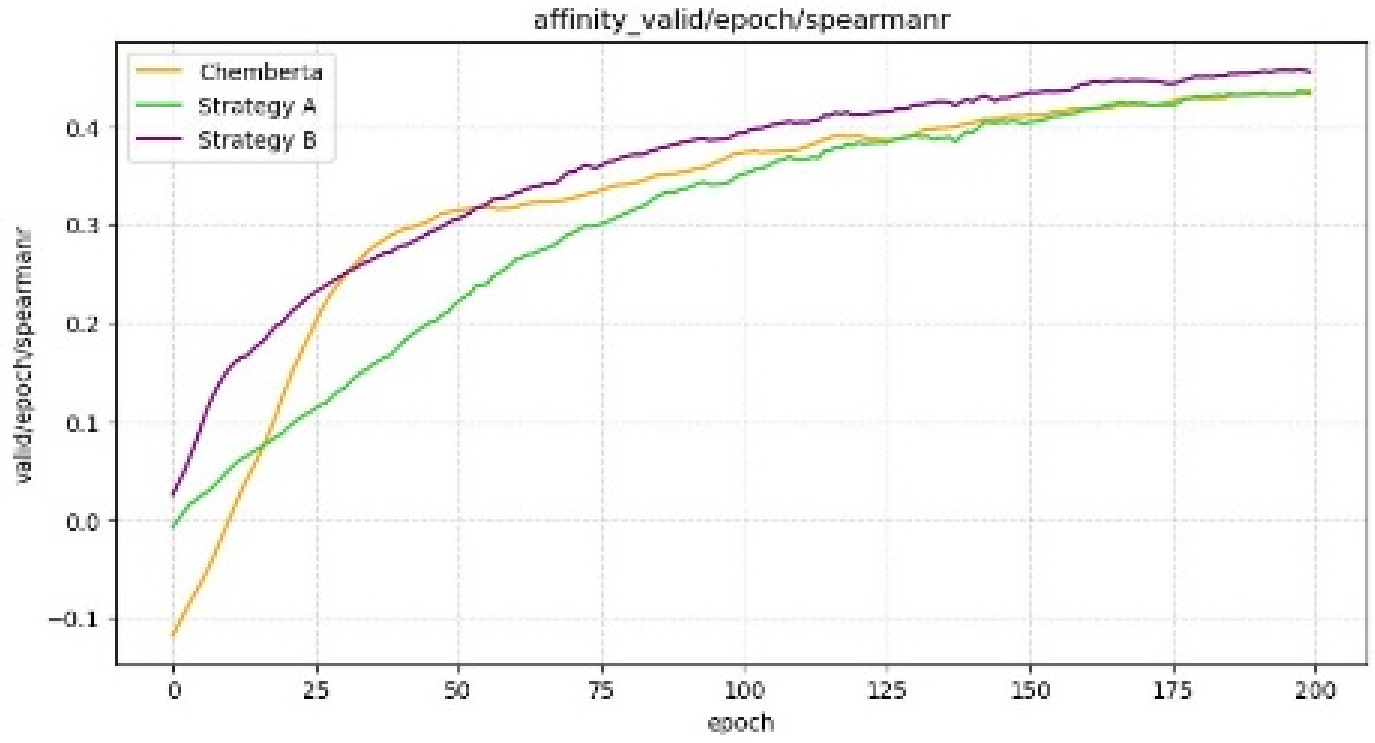

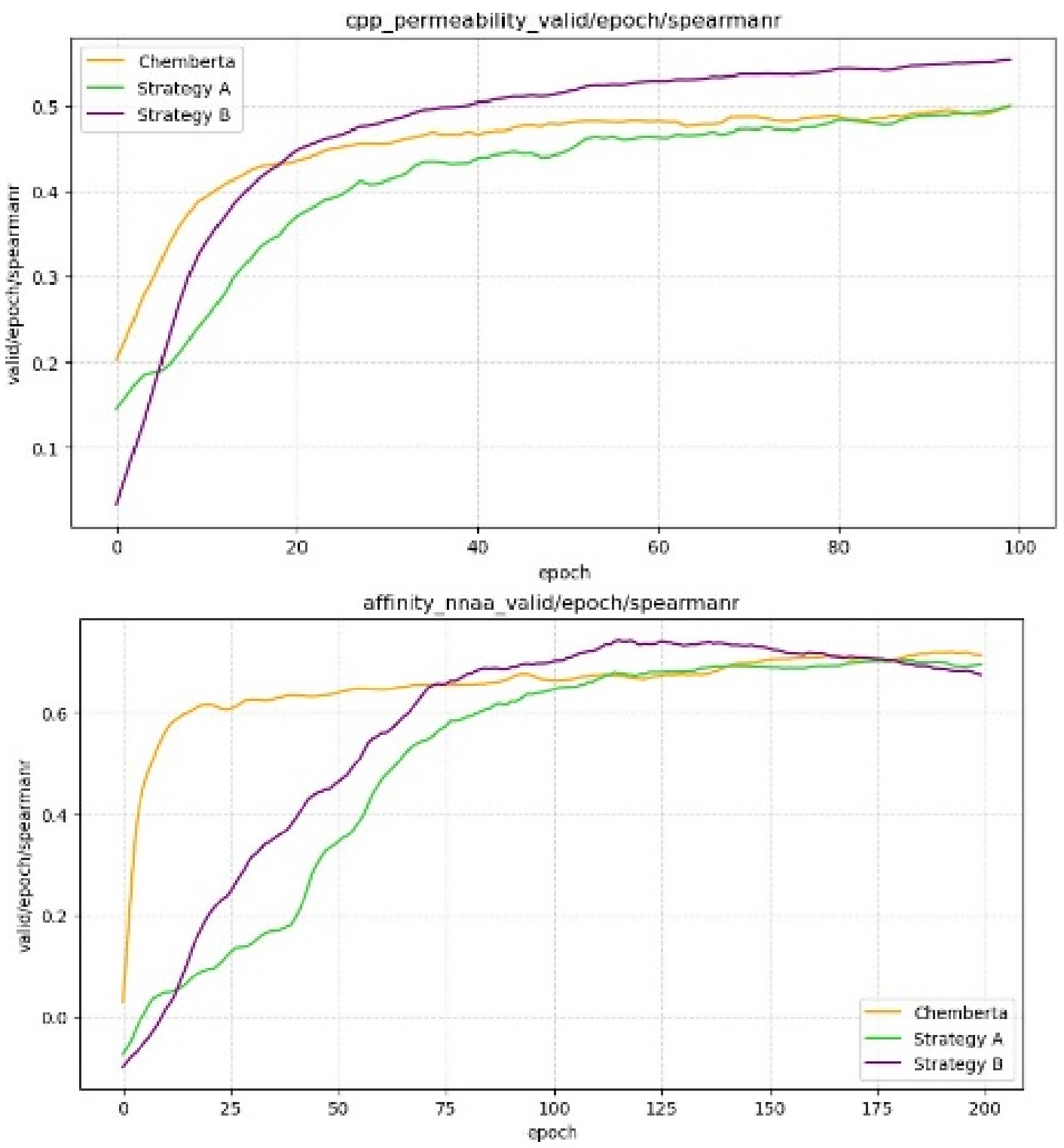

**Supplementary Figure S1.** The figure demonstrates the impact of the fragment splitting strategy on downstream tasks.

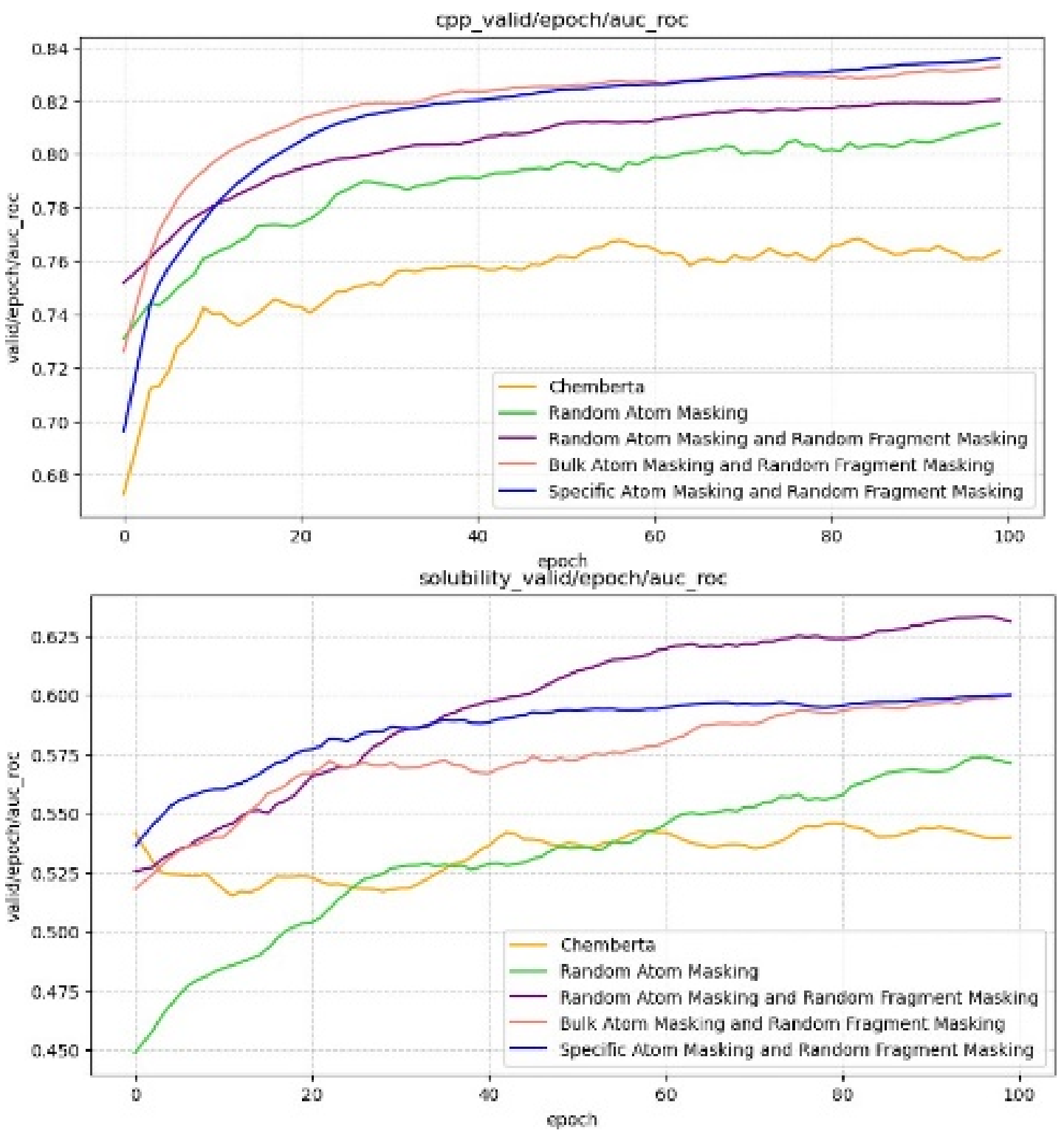

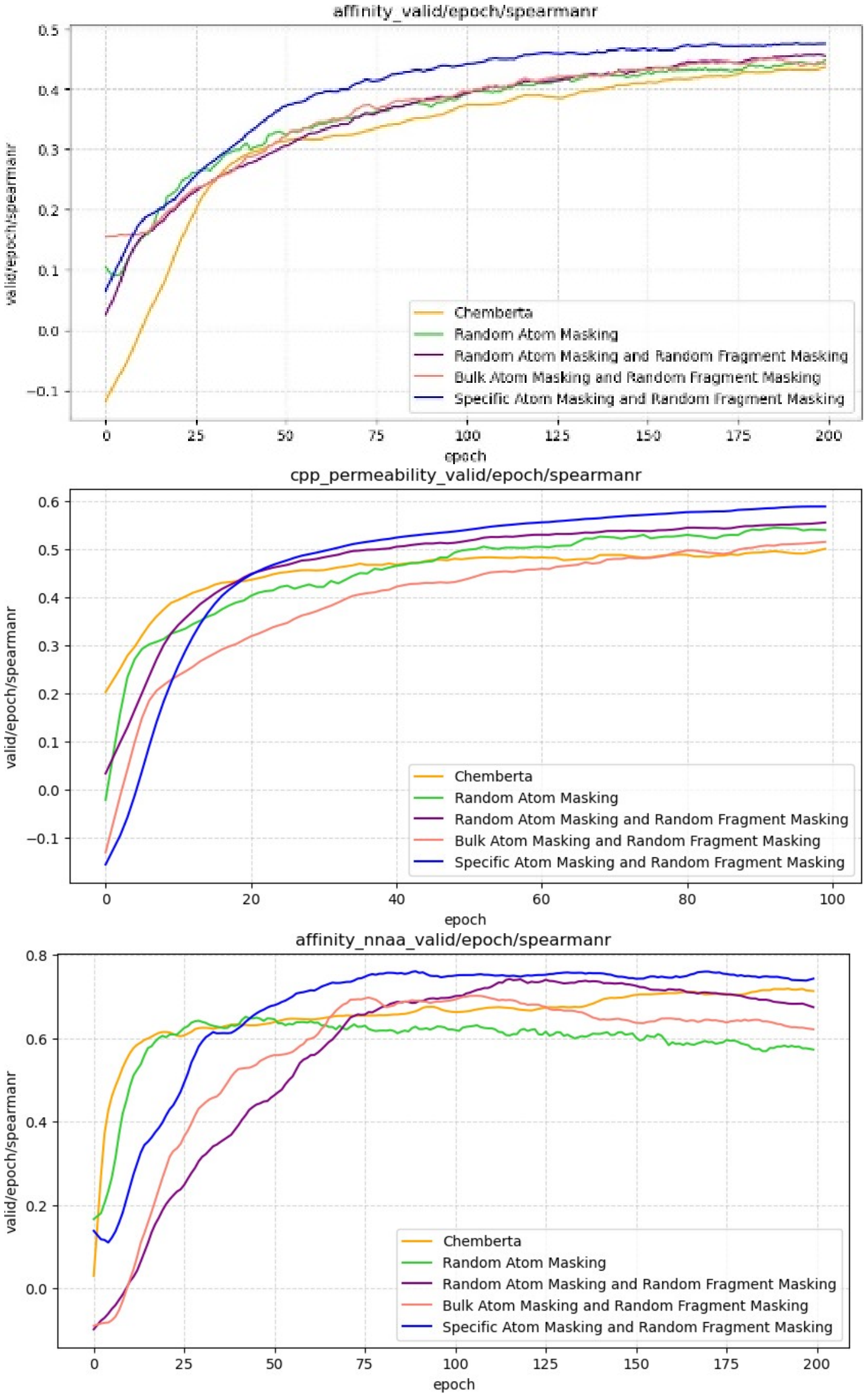

**Supplementary Figure S2.** The figure illustrates the influence of the masking method in pre-training models on downstream tasks.

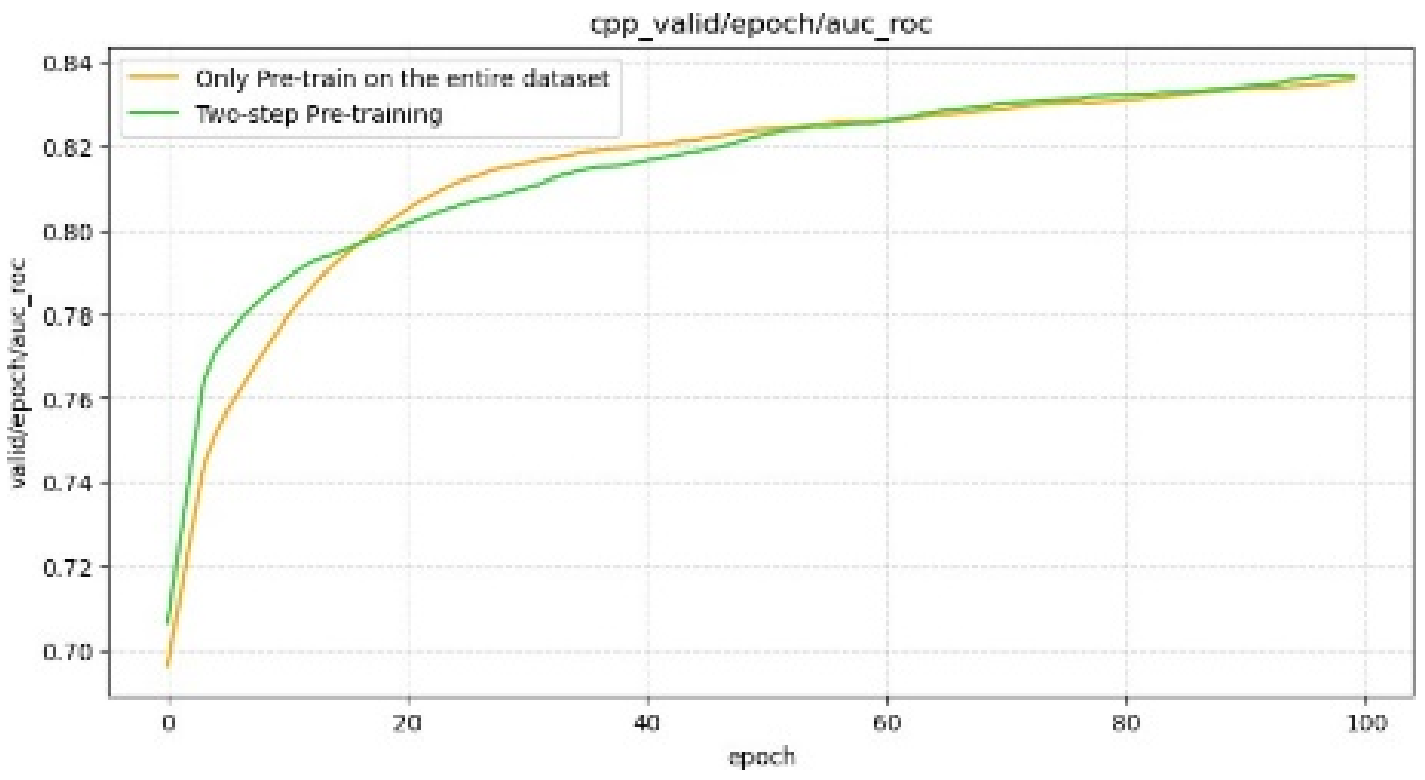

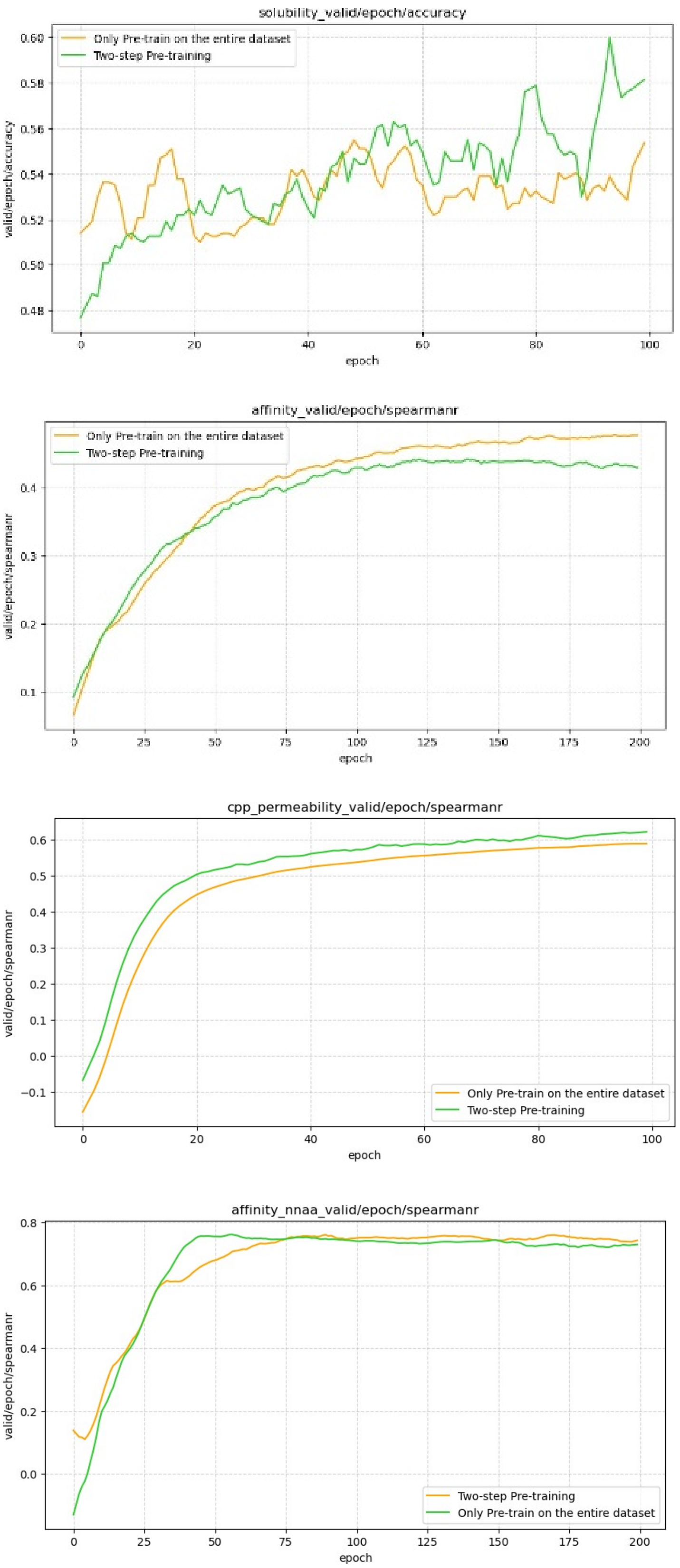

**Supplementary Figure S3.** The figure presents the impact of the two-stage pre-training on downstream

tasks.

# Supplementary Note 2: Hyperparameter Study

This section provides supplementary visualizations of fragment embedding, as depicted in the Figure 9.

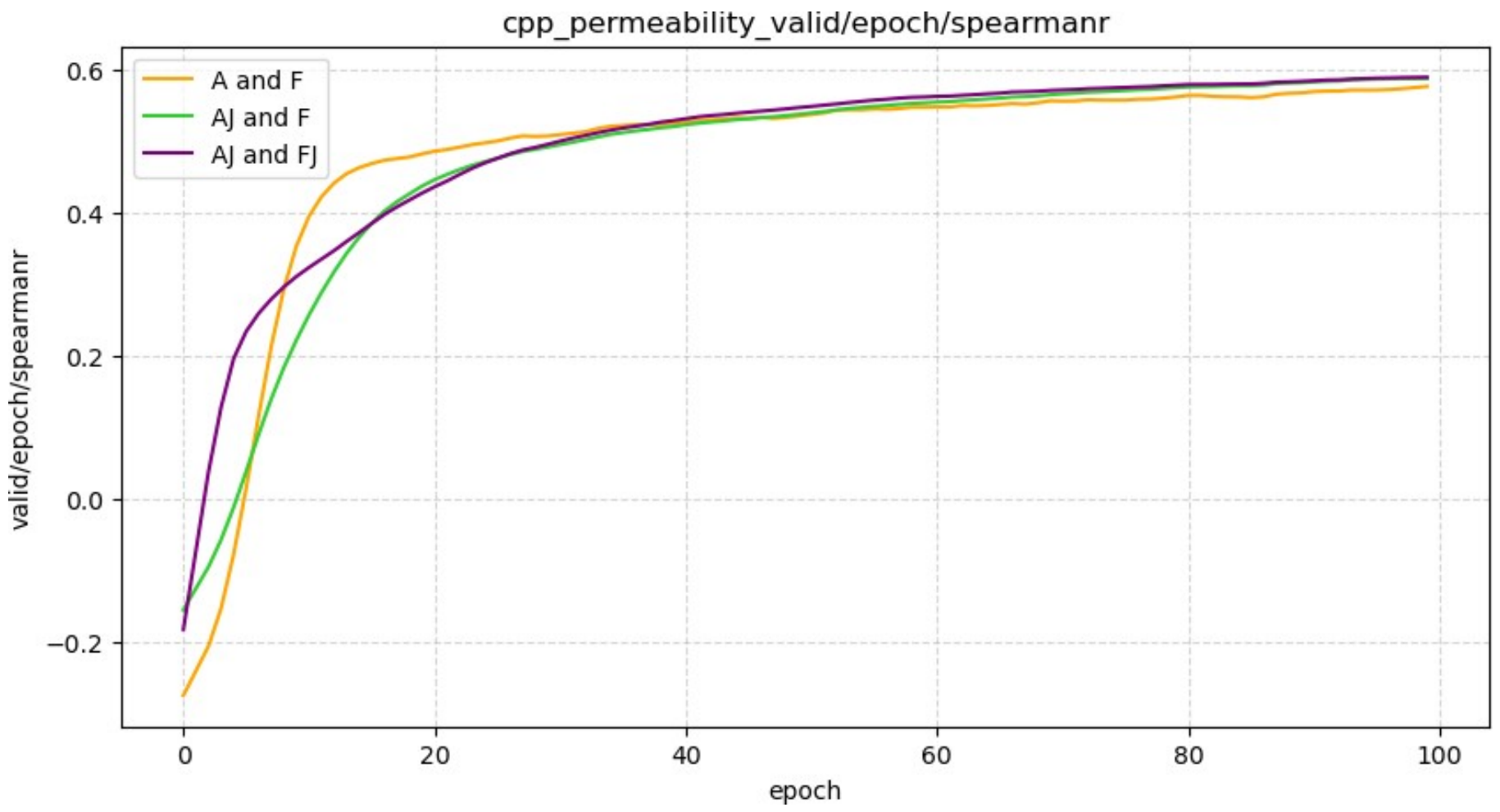

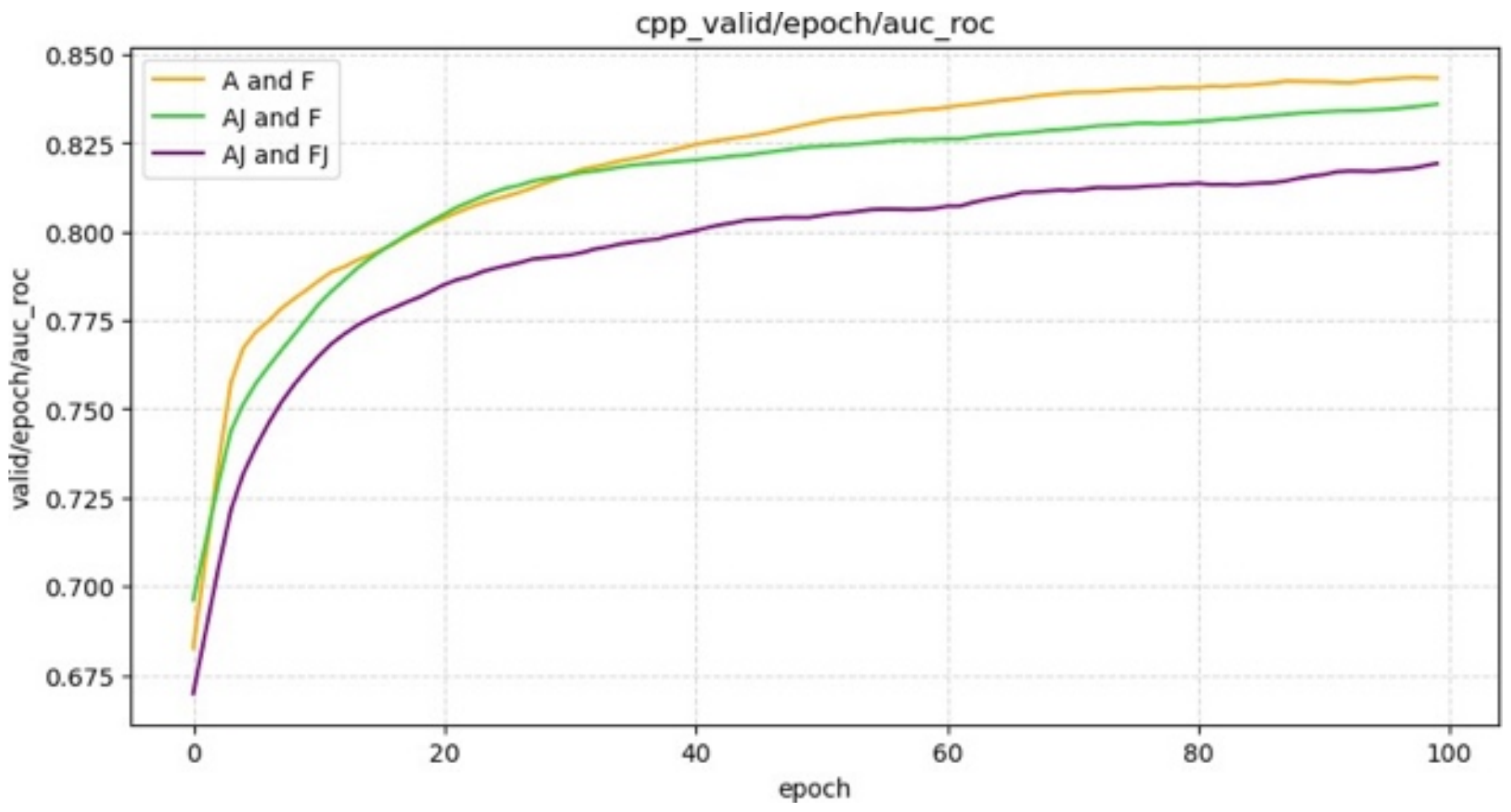

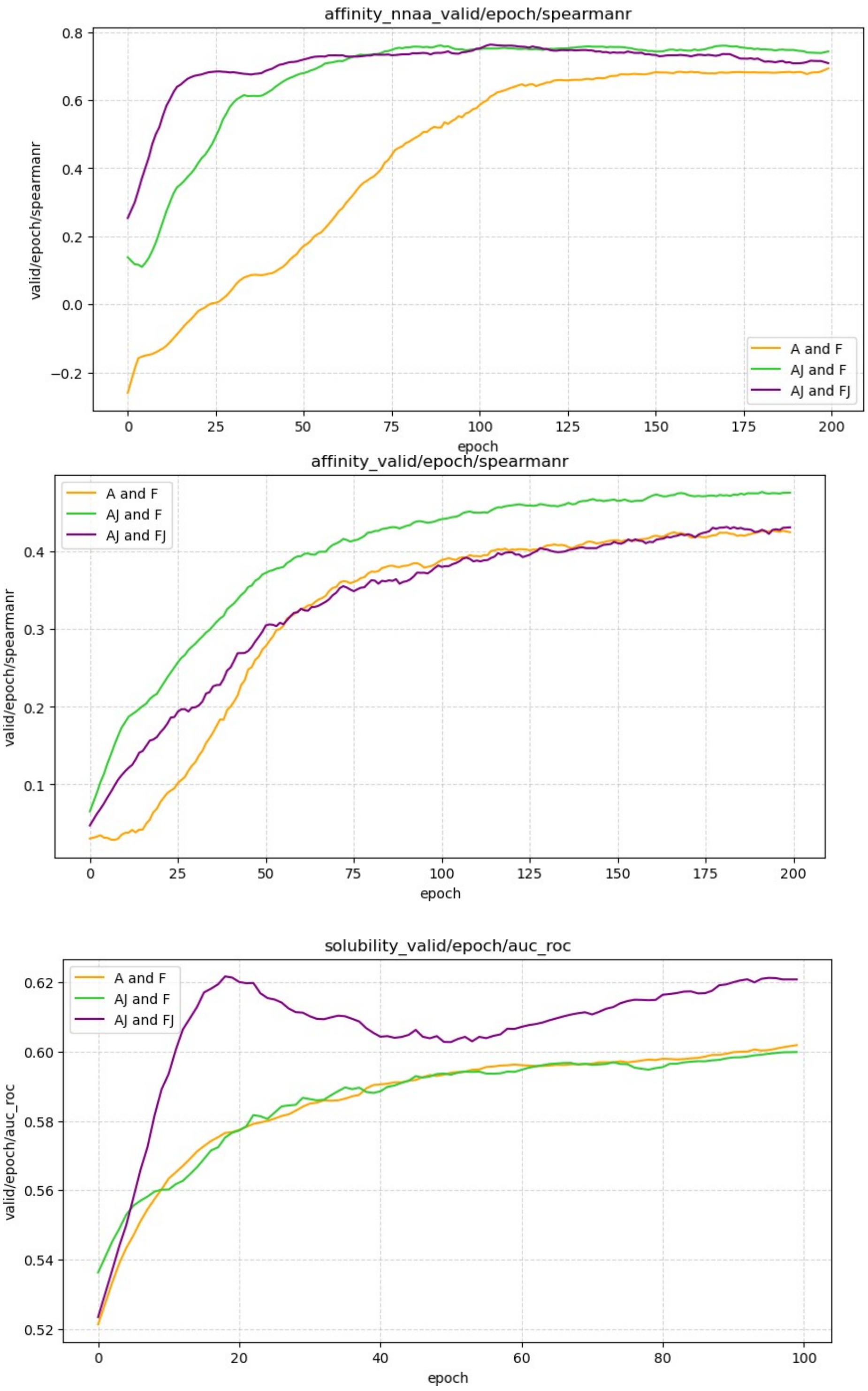

**Supplementary Figure S4.** The figure illustrates the selection of multi-view features in the model, where A represents atom-level features, J represents junction-level features, and F represents fragment-level features.